\documentclass{article}%
\usepackage{amsmath}
\usepackage{amsfonts}
\usepackage{amssymb}
\usepackage{graphicx}%
\providecommand{\U}[1]{\protect\rule{.1in}{.1in}}
\newtheorem{theorem}{Theorem}

\begin{document}

\title{Bondi-Metzner-Sachs symmetry, holography on null-surfaces and area
proportionality of \textquotedblleft light-slice" entropy\\{\small Dedicated to Detlev Buchholz on the occasion of his 65$^{th}$
birthday}\\to appear in Foundations of Physics}
\author{Bert Schroer\\CBPF, Rua Dr. Xavier Sigaud 150 \\22290-180 Rio de Janeiro, Brazil\\and Institut fuer Theoretische Physik der FU Berlin, Germany}
\date{May 2009}
\maketitle

\begin{abstract}
It is shown that certain kinds of behavior, which hitherto were expected to be
characteristic for classical gravity and quantum field theory in curved
spacetime, as the infinite dimensional Bondi-Metzner-Sachs symmetry,
holography on event horizons and an area proportionality of entropy, have in
fact an unnoticed presence in Minkowski QFT.

This casts new light on the fundamental question whether the volume
propotionality of heat bath entropy and the (logarithmically corrected)
dimensionless area law obeyed by localization-induced thermal behavior are
different geometric parametrizations which share a common primordeal algebraic
origin. Strong arguments are presented that these two different thermal
manifestations can be directly related, this is in fact the main aim of this paper.

It will be demonstrated that QFT beyond the Lagrangian quantization setting
receives crucial new impulses from holography onto horizons.

The present paper is part of a project aimed at elucidating the enormous
physical range of "modular localization". The latter does not only extend from
standard Hamitonian heat bath thermal states to thermal aspects of causal- or
event- horizons addressed in this paper. It also includes the recent
understanding of the crossing property of formfactors whose intriguing
similarity with thermal properties was, although sometimes noticed, only
sufficiently understood in the modular llocalization setting.

\end{abstract}

\section{Bondi-Metzner-Sachs symmetry, holography on null-surfaces and area
proportionality of "light-slice" entropy}

It has been known for a long time that the restriction of the vacuum state (or
any other finite energy state) to localized quantum matter results in a
thermal KMS state associated with a Hamiltonian which is uniquely associated
with these data \cite{Bi-Wi}. This knowledge has been mainly confined to a few
theoreticians with a foundational knowledge of QFT (local quantum physics
(LQP)). In the case where the localization behind a causal horizon is not a
Gedanken-construction (as e.g. in the Unruh Gedankenexperiment), but is
objectively fixed in form of an event horizon at a specific "place" in the
spacetime metric, it has attracted general attention. The best known example
is the case of a Hartle-Hawking state on global quantum matter in an extended
Kruskal-Schwarzschild world restricted to the outside of a Schwarzschild black
hole with the Hamiltonian describing the timelike Killing movement; in that
case the thermal aspect in form of the Hawking radiation became part of
popular knowledge.

It is perhaps less known that this \textit{localization-caused thermal
behavior} is also accompanied by a \textit{localization entropy}
\cite{interface} which behaves differently from the standard heat bath
entropy. In the case of (inside or outside) black hole localization this
entropy has nothing to do with the still illusive \textit{Quantum Gravity (QG)
}but belongs to the same thermal aspects as those discovered by Hawking.

The Localization-caused thermalization is methodologically related to the
conceptual setting of "holographic projection" in which a bulk algebra is
simplified by "projecting" it onto its causal horizon\footnote{The quoted
article is the most recent in a series of previous publications \cite{S1} in
which the modular localization formalism developed with the purpose to obtain
a conceptually sustainable basis of holography. In order to avoid any
confusion it should be mentioned that the rigorous quantum field theoretic
holography in this paper is a genuine \textit{projection }(reduction of
degrees of freedom). The reader should be aware that there is also a more
metaphoric use in discussions about quantum gravity where holography is
thought to lead to an \textit{isomorphic} storage of the information in the
bulk matter onto a "screen".}. Spacetime localization of quantum matter with a
sharp boundary causes infinitely large contributions from the vacuum
fluctuation at the causal boundary (horizon); allowing an appropriately
canonically defined "fuzzy" boundary in form of a "light slice" of thickness
$\Delta R$ leads to a logarithmically corrected area law for entropy and
energy. The physically relevant area is a dimensionless quantity and consists
of the (dimensionfull) geometric area of the horizon divided by the square of
the slice size $\left(  \Delta R\right)  ^{2}$ i.e. $a=\frac{Area}{(\Delta
R)^{2}}.$ It is modified for $\Delta R\rightarrow0$ by a factor $ln\alpha$.
This law is the same for all quantum matter, apart from a possible
contribution of the unknown quantum gravity, however as in the heat bath case,
a numerical factor c in front does depend on the kind of holographically
projected quantum matter.

The logarithmic correction which comes from the lightlike direction of the
"horizon-slab" may appear as an innocuous modification of the area law,
however it is essential for our understanding of the unity (the common root)
between heat bath and localization-caused thermal manifestation. Although this
localization entropy is not the primary topic of this paper, for reasons of
completeness we briefly present its derivation and add some new comments
(section 1.3).

The occurrence of infinite vacuum polarization at sharp boundaries and their
control by "softening" the boundary goes back to the dawn of QFT\footnote{The
reader is reminded that Heisenberg became aware of the presence of vacuum
polarization when he discovered that it is impossible to define a finite
"partial charge " of a region Q(V) without "softening " the boundary.}, but
the thermal characterization of the restriction of the vacuum state to the
operator algebra of a causally complete subregion is a combined result of
black hole physics \cite{Haw} and, in a more abstract conceptual setting, the
application of modular operator theory to the QFT subalgebra of operators
localized in a wedge region \cite{Bi-Wi}.

In this way the modular theory exposed the inexorable but often overlooked
link between quantum field theoretic localization and thermal manifestations.
The first who realized a possible physically relevant connection between the
observations of QFT in the presence of event horizons and modular situations
in QFT was Sewell \cite{Sew}\cite{Su-Ve}\cite{GLRV}. The thermal aspects of
modular theory should not came as a surprise since physicists, who discovered
certain aspects of this theory independent of mathematicians\footnote{The
theory bears the name \textit{Tomita-Takesaki modular theory}. Although Tomita
discovered it, the theory would not have arrived at how we know it without
essential contributions by Takesaki \cite{Tak}.}, obtained their results while
studying the conceptual problems of open systems in quantum statistical
mechanics \cite{Haag}. The more recent quantum field theoretical use of
modular theory for \textit{modular localization} is a adaptation of the
Tomita-Takesaki modular theory of operator algebras to causal localization of
states and operator algebras \cite{Haag}.

It is important to view the black hole physics within curved spacetime QFT and
the thermal consequences of localization within flat space QFT from a unified
standpoint since on the one hand such a viewpoint takes away from black hole
physics that the mysterious appearance of thermal effects popping apparently
out of nowhere when passing from flat to curved spacetime. On the other hand
it reminds us that there are important aspects in standard QFT which, as a
result of our Lagrangian prejudices, we have not been aware of.

The fact that only the event horizons in black hole physics are objective
placed horizons, whereas the causal horizons of quantum matter in Minkowski
spacetime are somewhat subjective (observer-dependent) Gedanken-horizons, is
no counter-argument since Gedankenexperiments as that of Unruh often lead to
corrections in our way of thinking; in addition the constants which appear in
the leading entropy behavior for the sheet size $\Delta R\rightarrow0$ are
interesting characteristics of the entire model, similar to Virasoro's
constant in chiral theories \footnote{The divergence of the localization
entropy for $\Delta R\rightarrow0$ is the only divergence with an intrinsic
significance i.e. which cannot be subsumed unter operator-valued distributions
(as fields and their correlations) and renormalization.} and not just of an
individual observables within the model.

The modern development of \textit{modular localization} of states and its use
for classifying and constructing QFT models will play an important role in
this paper.

. Although Minkowski space QFT does not know anything about the gravitational
interaction strength, and therefore by itself cannot produce a Bekenstein like
entropy formula (in which the dimensionless ratio is achieved with the help of
the gravitational constant instead of $\Delta R)$, it nevertheless does lead
to drastic change from the volume proportionality of heat bath entropy to the
(logarithmically modified) area behavior of localization entropy. The
derivation of the formula for localization entropy resulting from vacuum
polarization near horizons in which light sheets play a prominent role, is the
subject of the fourth section. Its existence has apparently been overlooked as
a result of a prejudice claiming that QFT cannot lead to an area law because
this allegedly requires a thinning out of degrees of freedom which only a
future quantum gravity (QG) could possibly achieve.

This more speculative role of holographic projections within a future theory
of quantum gravity, where it is expected to encode the full information
(holographic isomorphism) in the causally related bulk into a projected
"null-screen" \cite{Ho}, has been the point of departure of many recent
publications\footnote{Ever since 't Hooft's seminal work in which the concept
of holography appeared for the first time \cite{Ho}, there has been an
abundance of speculative quantum gravity inspired papers about holography on
null-surfaces. But in none of these papers a formulation of holography
consistent with the rather restrictive setting of QFT was attempted which
answers the question of how the holographic projection of a concrete bulk
theory really looks like as a QFT. One of the reasons may be that the
important mathematical instrument, namely modular localization, remained
largely unknown. In this work, as in a series of before cited previous
contributions, I try to redress this situation.}; in conjunction with
Bekenstein's black hole entropy proposal it led to formulas for gravitational
entropy bounds \cite{Bousso}. We have nothing to contribute to this problem
which is part of the still elusive QG; our results concern localization-caused
thermal behavior of QFT in Minkowski- and curved- spacetime and therefore aims
at the entropic counterpart of the Hawking temperature.

Likewise the holography used in the present work is a rigorous property within
the general setting of QFT\footnote{In order to keep things simple, we
restrict the bulk regions to wedges or double cones in Minkowski spacetime and
leave the extension to curved spacetime for future work.} and there are two
different ways of implementing it; one which starts with pointlike bulk fields
and produces pointlike generators of the holographic projected quantum matter
through intermediate semi-local steps, and the other starting from
wedge-localized algebras with the local substructure on its causal horizon
being constructed by algebraic intersections \cite{S1}. Both methods coalesce
on observables which are local in the sense of the lightfront and only arise
from integer dimensional bulk variables.

Whereas the algebraic method has been presented in previous work , the more
recent pointlike field method is the more appropriate for the present purposes.

The intimate interrelation of modular localization, thermal aspects and
gravitational localization (localization of quantum matter in front or behind
event horizons) begs the question whether other observations which have been
attributed to gravitation are also supported (possibly with a different
interpretation) by local quantum physics. Since QFT is a very well researched
subject from the viewpoint of Lagrangian quantization, one suspects that the
chances of making such observations increase if, as in the previous case of
thermal manifestations, one moves beyond perturbation theory into the more
general setting of local quantum physics.

In this note it will be shown that the Bondi-Metzner-Sachs (BMS) group
\cite{Sachs}, which originated in classical General Relativity within the
setting of asymptotically flat models, is really the symmetry of the vacuum
state restricted to local quantum matter holographically projected onto a
lightlike horizon. Whether this lightlike surface is the horizon of a compact
(double cone) or semi-compact region (Rindler wedge) in the bulk, or the
lightlike boundary of the entire cosmos in the sense of the asymptotic
flatness assumption of BMS, does not affect the mathematics but only the
physical interpretation.

This section continues in the form of three subsections. The first is
concerned with the definition and the basic properties of lightfront
holography. The appearance of infinite dimensional symmetry groups, including
the BMS group as a consequence of \textit{the symmetry enhancement of the
holographic lightfront projection} is the theme of the second subsection. Both
algebras are local in their own right, but somewhat nonlocal (semilocal in a
well-defined sense) relative to each other. The issue of localization entropy
is the subject of the third section. Here the knowledge of the lightfront
algebra is not sufficient because it encodes only the lightlike vacuum
polarizations within the lightfront and not those extending into ambient
spacetime. This requires to consider a light-slice of thickness $\Delta R$ and
to compute the leading power of the sheet entropy in the limit $\Delta
R\rightarrow0.$ The similarity to the thermodynamical limit turns out to be
more than an analogy. A precise definition of the light-slice entropy requires
the notion of \textit{modular localization} and the related \textit{split
property}.

Whereas the first and third subsection are extensions of already published
results \cite{S1}\cite{interface}, the derivation of the BMS symmetry from
null-space holography is new.

\subsection{ Holography on null-surfaces and the absence of transverse vacuum
polarizations}

As emphasized in the introduction, the only kind of holography which features
in this paper is the one which permits a rigorous formulation in QFT. The most
prominent case is the \textit{lightfront (LF) holography} which is essentially
the old \textit{lightfront quantization} but now with a more careful
formulation of what in the old days has been always neglected, namely the
relation with the original bulk description\footnote{The lightfront
quantization was mainly a computational device; how computed results connect
with precise observables in the bulk theory was usually left open.}. From the
point of symmetries, the restriction of the global bulk to the LF leaves a
7-parametric subgroup of the 10-parametric Poincar\'{e} group of 4-dimensional
Minkowski spacetime: 5 parameters account for a lightlike translation, a
lightlike dilation (the wedge-preserving boost transformation projected onto
the LF) and the 3-parametric transverse Euclidean group, whereas the remaining
two parameters are less obvious since they are the 2 "translations" of the
Wigner little group (3 dimensional Euclidean subgroup of the 6 parametric
Lorentzgroup) which leaves the lightray invariant \cite{S1}\cite{interface}.

If one uses characteristic data on the LF and propagates them into the bulk,
then the symmetry off the LF (i.e. the LF changing Poincar\'{e}
transformations) is certainly not encoded in the LF data; the first indication
that the bulk data can only be fully reconstructed from those which are
intrinsic to LF with extrinsic additional information.

The relation between LF characteristic data and bulk data is somewhat
nonlocal, compactly localized data on LF do not correspond to compactly
localized data in the bulk. A more informative way is to use the term
\textit{semilocal }since there is\textit{ a }global causal correspondence
between a wedge\footnote{As a standard wedge one usually takes the t-z wedge
$W=\left\{  \left\vert t\right\vert <z;\ (x,y)\in\mathbb{R}^{2}\right\}  ,$
all other wedges are Poincar\'{e} transforms. There is one Lorentz boost group
$\Lambda_{W}$ which leaves W$~$invariant and there is one reflection $j_{W}$
which maps $W$ into its causal opposite $W^{\prime}.$} $W$ and its (upper)
horizon $H(W)\subset LF(W)$ where the associated lightfront results from the
linear extension of $H(W)$. In classical (massive or massless) wave
propagation the symplectic subspace of $H(W)$ (characteristic) data on $H(W)$
corresponds to the symplectic subspace of $W-$ localized data. Obviously there
are many proper subregions (i.e. regions whose causal completion is not the
global spacetime) in the bulk whose horizon is on $LF$ consist of all wedges
$W\subset\mathfrak{W}$ (the set of all wedges) with $H(W)$ $\subset LF;$ they
are distinguished by the position of the $edge(W)\subset LF$ of their wedge.

A refinement of localization can be obtained by the formation of relative
causal complements. For example by translating a wedge by $a_{+}\ $along its
upper lightlike direction into itself $W\rightarrow$ $W_{a_{+}}\ $and forming
the relative causal complement%
\begin{equation}
W_{a_{+}}^{\prime}\cap W\equiv H(0,a_{+})
\end{equation}
where our notation indicates that the resulting region is a transverse
unbounded subspace on $H(W)$ of lightlike extension $(0.a_{+}).$ Since all
$W^{\prime}s$ result from a fixed one by Poincar\'{e} transformations, this
relative causal complement construction can be generalized to the other $LF$
preserving transformations. It turns out that the relative causal complements
define a local structure on $LF$ which contains arbitrary small regions.
However this does not work in the other direction: from the local data on LF
one can only reconstruct the data for the LF compatible W's and as we saw,
their relative commutants do not lead to new proper bulk regions. This means
that we are unable to construct the bulk substructure e.g. that of double
cones $D\subset W$ inside $W.$ The group theoretic reason is that the
holograpically projected lightfront world is (as the name projection suggests)
is not isomorphic to the Minkowski world since the holographical symmetry
group is a proper subgroup. If on the other hand the symmetry groups are
shared as in the famous AdS$_{5}$-CFT$_{4}$ correspondence\footnote{This is
special case of a correspondence between certain regions of two manifolds
which, although different by one dimension, share their maximal spacetime
symmetry group. The correspondence extends to the operator algebras associated
with those region, but its physical. Its physical usefulness is however
doubtful \cite{Sw}.}, the projection may change into an isomorphism. However
even in this case there is no invertible relation between generating pointlike
quantum fields. In the present context holography will always refer to a
horizon and all horizons are null-surfaces.

After having explained the kinematical prerequisites of lightfront holography
one can now fill it with a dynamical content. In the classical setting of
massive or massless linear wave theories when the full space is a symplectic
space and the space of waves localized in the causal complement is identical
to the symplectic complement of the subspace associated with the original
localization. In the case of local quantum physics the local observables form
operator algebras which act in a Hilbert space. The vacuum representation
consists of all operators in a Hilbert space $B(H)$, the commutants of
$\mathcal{O}$-localized subalgebras $\mathcal{A(O)}^{\prime}\subset B(H)$
replace the classical symplectic complements and the Haag duality relation
$A(\mathcal{O})^{\prime}=A(\mathcal{O}^{\prime})$ extends the classical
analogy. The nontriviality of the lightfront holography corresponds to the
nontriviality of intersections within the family of wedge algebras whose
horizon lies on the same LF. is nontrivial if the intersections are not the
trivial algebra $\left\{  \lambda\mathbf{1}\right\}  .$

The value of the holographic projection is a significant simplification of the
dynamical problem at the expense of the loss of some information about the
bulk. The quantum origin of this simplification is very interesting since it
consists in the absence of transverse vacuum polarization; in other words
\textit{the vacuum tensor factorizes in transverse direction} (see below) so
that the vacuum polarization is limited to the lightray direction and hence
there is \textit{no lightlike tensor factorization} of the vacuum. This kind
of conceptual preparation helps to avoid making interpretational mistakes in
attempting to define holographic projections directly in terms of pointlike
fields. For our main purpose, namely the derivation of the BMS group from the
symmetry enhancement of the holographic projection, we only need the latter in
the absence of interactions.

The crucial property which permits a direct holographic projection for a free
field is the \textit{mass shell representation} of a free scalar field%
\begin{equation}
A(x)=\frac{1}{\left(  2\pi\right)  ^{\frac{3}{2}}}\int(e^{ipx}a^{\ast}%
(p)\frac{d^{3}p}{2p_{0}}+h.c.) \label{mass}%
\end{equation}
Using this representation one can directly pass to the lightfront by using
lightfront adapted coordinates \ $x_{\pm}=x^{0}\pm x^{3},~\mathbf{x},$ in
which the lightfront limit $x_{-}=0$ can be taken without causing a divergence
in the p-integration. Using a p-parametrization in terms of the wedge-related
hyperbolic angle $\theta:p_{\pm}=p^{0}+p^{3}\simeq e^{\mp\theta},~\mathbf{p}$
the $x_{-}=0$ restriction of $A(x)\ $to LF reads \cite{S1}%

\begin{align}
&  A_{LF}(x_{+},\mathbf{x})\simeq\int\left(  e^{i(p_{-}(\theta)x_{+}%
+i\mathbf{px}}a^{\ast}(p_{-},\mathbf{p})d\mathbf{p}\frac{dp_{-}}{2p_{-}%
}+h.c.\right)  \label{LF}\\
&  \left\langle A_{LF}(x_{+},\mathbf{x}),A_{LF}(x_{+}^{\prime},\mathbf{x}%
^{\prime})\right\rangle \simeq\delta(\mathbf{x}-\mathbf{x}^{\prime})\frac
{1}{2\pi}\int e^{-ip_{-}(x_{+}-x_{+}^{\prime})}\frac{dp_{-}}{2p_{-}%
}\nonumber\\
&  \left[  A_{LF}(x_{+},\mathbf{x}),A_{LF}(x_{+}^{\prime},\mathbf{x}^{\prime
})\right]  \simeq\delta(\mathbf{x}-\mathbf{x}^{\prime})\frac{1}{2\pi}%
\int(e^{-ip_{-}(x_{+}-x_{+}^{\prime})}-e^{-ip_{-}(x_{+}^{\prime}-x_{+})}%
)\frac{dp_{-}}{2p_{-}}\nonumber
\end{align}
Different from the bulk field, these formula for the two-point function on LF
contain a logarithmic infrared divergence if interpreted pointwise. To see
that this is harmless one recalls that fields are singular objects
(operator-valued distributions) which only lead to observables after smearing
with test functions. As a result of the mass shell restriction the equivalence
class of test function which have the same restriction $\tilde{f}|_{H_{m}%
}=P_{m}f$ to the mass hyperboloid of mass $m$ is mapped to a unique test
function $f_{LF}$ on the lightfront \cite{Dries}\cite{S1}%
\begin{equation}
A(f)=A(\left\{  f\right\}  )=A_{LF}(f_{LF})
\end{equation}
However the image on LF does not contain all Schwartz test
functions\footnote{The testfunctions ("smearing"-functions) in Wightman QFT
are members of the Schwartz testfunctions space i.e. smooth functions of fast
decrease. Although their Fourier transforms have the same properties, their
restriction to the mass shell and the subsequent transformation to the
lightfront coordinates leads to a subspace of Schwartz functions which is
compatible with the logarithmic behavior and the positivity of the LF
two-pointfunction.} but only such functions whose total x$_{-}$ integral
vanishes i.e. which do not "see" the logarithmic infrared divergence at
$p_{-}=0$ of the pointlike formula (\ref{LF}). This problem and its cure via
testfunction restriction is well-known from the zero mass scalar field in
d=1+1 and can be traced back to the fact that this field in lightlike
direction is really a semiinfinite integral over a chiral current, which makes
it string-localized\footnote{the $\partial_{+}$ derivative of $A_{LF}$ which
is a pointlike field and can be smeared with unrestricted Schwartz test
function serves as a better generator of the LF algebra. Derivatives of local
fields generate (after Haag dualization) the same localized algebras
$\mathcal{A(O})$ as the field.}. The simplifying feature of $A_{LF}%
(x_{+},\mathbf{x}\dot{)}$ is, as expected according to the previous
considerations, the absence of transverse vacuum fluctuations as evidenced by
the appearance of quantum mechanical $\delta(\mathbf{x-x}^{\prime})$ delta
function. In fact the LF generator behaves as a transverse extended chiral
theory\footnote{The derivative $\partial_{x_{+}}A_{LF}$ defines a transverse
extended (direct integral) abelian chiral current operator.} of a chiral field
which is the potential associated to a chiral current field. The vacuum
polarization phenomenon, which is closely related to the energy positivity in
conjunction with a finite propagation speed, has moved exclusively into the
lightlike direction whereas in the transverse direction the vacuum factorizes.
Holography on null-surfaces is the only construction in QFT which leads to a
partial tensor factorization as a result of directional absence of vacuum polarization.

Although there is no local relation between the data on LF and those in the
bulk, there is a remnant of "semi"-locality between the data on that part of
LF which coincides with the horizon the wedge $H(W)$ and the bulk algebra on
$W;$ instead of the equality of the associated classical subspaces one now has
an equality of operator subalgebras of the algebra of all operators $B(H).$
The meaning of the statement that the field $A_{LF}(\mathbf{x,}x_{+})$ is a
generator of the bulk algebra $\mathcal{A}(W)$ becomes now clear since the
equality between individual operators of smeared fields $A(f)|_{suppf\in
W}=A_{LF}(f_{LF})$ can be lifted to the equality of the operator
algebras\footnote{It is very important not to misunderstand the notation
$\mathcal{A}(W)\subset B(H);$ it does \textit{not} imply any knowledge about
the substructure about the local net inside $W.$ The local substructures of
$\mathcal{A}(W)$ and $\mathcal{A}(H(W))$ are of course very different (
geometric subregions in $H(W)$ have no geometric counterpart in $W$ ), only
the position of these global algebras inside B(H) is identical.}
\begin{align}
\mathcal{A}(W)  &  =\mathcal{A}(H(W)),~H(W)\subset LF(W)\label{hol}\\
\mathcal{A}(W)  &  \equiv alg\left\{  e^{iA(f)}|suppf\subset W\right\}
\nonumber
\end{align}
where the last line denotes the (weakly closed) operator algebra generated by
the Weyl elements which are exponentials of free Bose fields; in case of free
Ferimons the CCR (Weyl) algebra is replaced by the CAR algebra. The use of
Weyl algebras in QM is probably well-known to readers who worked with coherent
states. The corresponding LF algebras $\mathcal{A}(H(W))$are defined
correspondingly with supp$f_{LW}\subset H(W)\subset LF$ leading to analytic
functions on a strip in the appropriate momentum space rapidity which obey the
mentioned vanishing of their $x_{+}~$Fourier transform at $p_{-}=0.$

We are now getting to a conceptual subtlety of lightfront holography. In
writing such relations between bulk and horizon algebras as above, the
algebras are meant without any knowledge of their local substructure e.g.
knowing $\mathcal{A}(W)$ only means that its relative position (inclusion) in
$B(H)$ is known, but not the local substructure of $\mathcal{A}(W)$ i.e. the
position of sharper localized algebras inside $\mathcal{A}(W).$ Knowing a
generating pointlike field for the bulk (\ref{mass}) one of course knows the
position of all operator algebras for arbitrary causally closed bulk regions,
but in writing algebraic relations (\ref{hol}) we have to forget about the
local substructure. In particular the knowledge of $\mathcal{A}(W)$ should not
be confused with knowing the pointlike generators in the region W.

A localized algebra is a holistic object in whose definition the localized
test functions of individual elements which entered its construction are lost;
sharper localized subalgebras can only be regained by either appropriately
intersecting wedge algebras or going back to the generating bulk fields and
repeating similar constructions for sharper localized test function. In
particular $\mathcal{A}_{LF}$ contains only lightlike vacuum fluctuations. The
physical questions concerning localization entropy are however not related
with vacuum polarization \textit{on} the horizon but rather \textit{near} the
horizon and for their study the knowledge of $\mathcal{A}(W)\subset B(H)$ is
insufficient since it says nothing about the local substructure. In the third
subsection we will address the localization entropy in a thin light sheet;
this requires different techniques that those of this and the next section.

The knowledge of $\mathcal{A}(W)$ but without its substructure is a state of
affairs which one can only describe in terms of the position of an operator
subalgebras within $B(H),$ but it is not possible to encode this partial
knowledge in terms of fields. Therefore it is not surprising that in a
constructive approach where it is essential to dissect the difficult problem
of constructing interacting fields (or, what is the same, the complete theory)
into simpler parts and solve one problem at a time, operator algebras provide
an important tool

In connection with the AdS$_{n+1}$-CFT$_{n}$ correspondence it was mentioned
that in cases of relations between theories in different spacetime dimensions
an encoding into pointlike fields has to be replaced by one between algebras
of certain (usually noncompact wedge-like) regions. In order to arrive at
arbitrarily small double cone algebras (the closest approximation to pointlike
fields) one has to study intersections of those algebras which one obtained
from the correspondence or holography, but the latter does not extend to these
smaller algebras.

This can also be seen in terms of the LF fields. The fields on $H(W)\subset
LF$ are subject to a significantly different localization from that of the
bulk $W.$ In particular it is not possible to reconstruct the local
substructure of $\mathcal{A}(W)$ even though globally $\mathcal{A}%
(W)=\mathcal{A}(H(W)).$ The field $A_{LF}(x_{+}\mathbf{x})$ does not know
anything about the physical mass, but with its value added one can compute
$\mathcal{A}(W)$ with its bulk substructure. The relevant propagation formula
from the null-surfaces into the bulk is contained in a recent paper \cite{Re}%

\begin{equation}
A_{m}(x)=-2i\int_{LF}dy_{+}d^{2}y_{\perp}\Delta_{m}(x-y)|_{y_{-}=0}%
A(y_{+},y_{\perp})\label{com}%
\end{equation}
where $\Delta_{m}$ is the massive commutator function and $A(y_{+},y_{\perp})$
is the field on the lightfront. If one wants to compute the $A_{m}(x)$ for
$x\in W$ we only have to integrate over the horizon $y\in H(W)\subset LF$ and
not over all of LF; this is automatically incorporated in (\ref{com}) via the
support properties of $\Delta_{m}.$ Note that the holographic data, in
distinction to Cauchy data, do not know anything about the mass which has to
be added from the outside. There are of course other, philosophically more
challenging ways to supply the missing information. Their common feature is to
get back from the 7 parametric subgroup to the full 10-parametric Poincar\'{e}
group. A particular interesting way is to do this via a GPS-like positioning
method using several (in d=1+3 one needs maximally 3) holographic projections
(in certain relative modular positions \cite{interface}); in this case the
mass enters indirectly via the 10-parametric Poincar\'{e} group which in turn
result from 7-parametric subgroup in different LF positions. Hence the
concreteness of pointlike holographically projected fields is somewhat of an
illusion since in general there is no direct pointwise relation between them
and the bulk; what is really identical objects are certain noncompact algebras
which they generate and which by intersections lead to the local bulk structure.

The null-surface propagation formula can also be used to demonstrate that this
kind of inverse holography is capable to solve important problems which other
methods, as e.g. functional analytic methods, failed to do. It is well-known
that the modular group of $\mathcal{A}(W)$ acts on the generating field
$A_{m}(x)$ as the $W$-preserving Lorentzgroup, whereas on the generator
$A_{LF}(x_{+},\mathbf{x})$ of $\mathcal{A}(H(W))$ it acts as the $x_{+}$
dilation. Clearly the action on the holographic projection is simpler. It is
now easy to see that the hologaphic inversion formula does map the dilation
group into the W-preserving Lorentz group. Since in (\ref{com}) the scale
transformation on $A(y_{+},y_{\perp})$ acts on $y_{+}$ and trivially on the
fixed point $y_{-}=0,$ we can shift the scale transformation to the second
argument of the commutator function. Using the Poincar\'{e} invariance of the
latter and renaming integration variables one can shift the scale
transformation $y_{+}\rightarrow\lambda^{-1}y_{+},y_{-}\rightarrow\lambda
y_{-}$ to the unintegrated bulk variable whereupon fit takes the form of a
W-preserving Lorentz group.$~$\ Note that whereas for the Cauchy propagation
every spatial region independent of its size has its nontrivial causal shadow
region, this is not the case for the inverse holography, in that case the
region must be of the form $H(W)\subset LF$ and for a given $LF$ there are of
course many $H(W)^{\prime}s.$

A similar formula to (\ref{com}) serves in order to extend inverse holography
for massive free fields to other horizons as those of (noncompact) spacelike
cones $H(\mathcal{C})$ or (compact) double cones $H(\mathcal{D})$. In that
case it is well known that the modular groups associated to the bulk
algebras\footnote{If not stated otherwise the knowledge of a local algebra
only means its positioning in the algebra of all operators in the Hilbert
space $\mathcal{A(O})\subset B(H)$ and does not include its net-structure
inside $\mathcal{O}.$} $\mathcal{A}(\mathcal{C}),~\mathcal{A(D}).$ act in a
"fuzzy" i.e. nongeometric way. Let us now see how the "fuzzyness" develops in
the case of a double cone. For simplicity we stay in d=1+1 and chose a double
cone symmetric around the origin as in \cite{Haag}. Then the lower mantle of
the cone with apex (-1,0) is a Horizon whose causal shadow covers the double
cone. Every signal which entered the double cone must have passed through the mantle.

In this case the propagation from the two pieces of the mantle leads to the
sum%
\begin{align}
&  A_{m}(x_{+},x_{-})=-2i\int_{-1}^{+1}dy_{+}\Delta_{m}(x-y)|_{y_{-}=0}%
A(y_{+})+\\
&  +-2i\int_{LF}dy_{-}\Delta_{m}(x-y)|_{y_{+}=0}A(y_{-})\nonumber
\end{align}
Now the modular group on the horizon acts fractional namely the "dilation"
which leaves the fixed points $y_{\pm}=-1,+1$ invariant (instead of $0,\infty$
as in the first case). The modular group on both parts of the horizon is%
\begin{equation}
x_{\pm}(s)=\frac{(1+x_{\pm})-e^{-s}(1-x_{\pm})}{(1+x_{\pm})+e^{-s}(1-x_{\pm})}%
\end{equation}
Different from the previous case, one cannot transfer this fractional change
from the $y$ to the $x$. There is no local fractional transformation on the
bulk, rather the action is fuzzy but stays inside the double cone as it
should. It is however not purely algebraic since it was obtained by combining
the geometric group on the horizon with the causal propagation whose
"reverberation" (non-Huygens) aspect causes the fuzzyness\footnote{By
construction the modular action restricted to the horizon is always geometric,
which is closely related with the modular preservation of horizons.}. In a
forthcoming paper it will be shown that the modular unitaries $\Delta
_{\mathcal{O}}^{it}$ for arbitrary localization regions do not depend on the
interaction i.e. can be computed in the incoming free field theory; all the
dynamical dependence is contained on the modular reflection $J_{\mathcal{O}}$.
A calculation of the pseudodifferential generators of these fuzzy modular
actions for the double cone algebra generated by free fields will be contained
in a forthcoming paper by Brunetti and Moretti\footnote{I thank Romeo Brunetti
for informing me forthcoming results.}.

The most important message one can learn from this simple setting of
holography is that contrary to popular belief one cannot encode the full
information about the theory in the bulk into a holographic screen; in the
above presentation we emphasized that the value of the mass must be added, but
one could also say the holographically projected correlation functions do not
allow to reconstruct the creation/annihilation operators $a(p)^{\#}$ or
equivalently that the action of the LF-changing remaining three Poincar\'{e}
transformation have to be supplied from the outside. 

The difference beween the old "lightcone quantization" and modern holography
is that the "lightcone fields" were not considered as differently localized
objects in the same theory, but rather as objects resulting from a different
quantization. This created an enormous amount of confusion over many decades.
Even though the free field holography is too simple to illustrate the
advantage of the method in any convincing way, it does reveal its main
philosophical basis: a radical change of the spacetime ordering device for a
given kind of quantum matter\footnote{In the free $d\geq4$ case there is
(according to the commutation structure) only CCR and CAR quantum matter.} in
which part of the original information (e.g. mass spectra, particle
properties) gets lost, but certain other structural properties permit a
simpler description.

Physics is not defined in terms of abstract pure algebraic structure, but
rather requires spacetime ordered quantum matter. Hence holography should be
conceptually placed together with the change of spacetime ordering in passing
between isometric submanifolds of different curved space time QFTs using the
same abstract quantum matter (see the local covariance principle of QFT in CST
\cite{Brun}). In both cases the work is still in its beginnings, but it is
already clear that the spacetime ordering aspect in those new ways of looking
at QFT is much more flexible than in the old "lighcone quantization" or in the
old way of treating QFT in CST for each fixed CST separately.

There is another remarkable observation about a gain in spacetime symmetry. Up
to now we only mentioned the \textit{symmetry loss} from the Poincar\'{e}
symmetry in the bulk to its 7-parametric LF subgroup, but there is also an
enormous symmetry-gain which has no counterpart in the bulk theory.
Transversely extended chiral theories share the infinite Diff(S$^{1}$)
invariance of chiral theories. In order to identify the LF projection with the
well-known theory of an abelian current we take the x$_{+}$ derivative
$\partial_{x_{+}}A$ and obtain%
\begin{equation}
\left\langle \partial_{x_{+}}A_{LF}(x_{+},\mathbf{x})\partial_{x_{+}}%
A_{LF}(x_{+}^{\prime},\mathbf{x}^{\prime})\right\rangle \simeq\delta
(\mathbf{x}-\mathbf{x}^{\prime})\frac{1}{\left(  x_{+}-x_{+}^{\prime
}-i\varepsilon\right)  ^{2}} \label{cur}%
\end{equation}
where the second factor describes the correlation of the chiral current whose
commutator is a $\delta^{\prime}$ function describes the of fluctuating chiral
current. Keeping the transverse coordinates fixed and only transforming the
compactified (always possible for chiral theories) lightlike coordinates, the
maximal symmetry is\footnote{The vacuum preserving symmetries consist only of
the Moebius subgroup.} Diff(S$^{1}$). This is a well-known consequence of
properties of chiral energy-momentum tensors. Actually the symmetry is even
larger because there are also transverse and mixed transvere-lightlike
symmetry transformations. We will return to this problem in the next section.

The success of the pointlike formulated lightfront holography in the absence
of interactions begs the question whether there exists a pointlike formulation
in the presence of interactions. The main problem in this case is to find an
analog of the mass shell representation (\ref{mass}). Such representations for
interacting fields already appeared in the 60s; shortly after the formulation
of LSZ scattering theory Glaser, Lehmann and Zimmermann introduced such
representations which became known as "GLZ representations" \cite{GLZ}. They
express the interacting Heisenberg field as a power series in incoming
(outgoing) free fields. In case there is only one type of particles one has:
\begin{align}
&  A(x)=%
%TCIMACRO{\dsum }%
%BeginExpansion
{\displaystyle\sum}
%EndExpansion
\frac{1}{n!}\int_{V_{m}^{\pm}}..\int_{V_{m}^{\pm}}a(p_{1},..p_{n})e^{i\sum
p_{k}x}:A_{in}(p_{1})..A_{in}(p_{n}):\frac{d^{3}p_{1}}{2p_{10}}..\frac
{d^{3}p_{n}}{2p_{n0}}\label{GLZ}\\
&  A_{in}(p)=a_{in}^{\ast}(p)~on~V_{m}^{+}~and~a_{in}(p)~on~V_{m}%
^{-}\nonumber\\
&  a(p_{1},...p_{n})_{p_{i}\in V_{m}^{+}}=\left\langle \Omega\left\vert
A(0)\right\vert p_{1},...p_{n}\right\rangle \text{ ~}%
vac.pol.component\nonumber
\end{align}
where the integration extends over the forward and backward mass shell
$V_{m}^{\pm}\subset V_{m}$ and the product is Wick ordered. Under the
assumption of asymptotic completeness the incoming fields form a complete set
and the GLZ formula (without a statement about its convergence status) is
nothing more than a formal way of encoding all matrix elements of the field
$A(x)$ between l ket and k bra incoming states with n=l+k. The coefficient
functions $a(p_{1},...p_{n})$ are mass shell restriction of retarded functions.

Even though there is no control about the convergence \footnote{In contrast to
the perturbative expansion which is known to diverge (even in the Borel
sense), the convergence status of GLZ had not been settled. In d=1+1
factorizing models there are some indications in favor of convergence.}, but
at leas superficially the formal lightfront restriction for each term in
(\ref{GLZ}) does not seem to cause short distance divergences\footnote{This is
strictly speaking only true for d=1+1 theories which have no transverse
degrees of freedom. For higher space-time dimension there are problems from
composite fields (see below).}. It is also clear that (as in the case of free
fields) it is not possible to define a lightfront restriction in terms of
vacuum expectations (Wightman functions). But within a mass shell
representation as (\ref{GLZ}) the limit can be formally taken term for term by
setting $x_{-}=0$ with the result%
\begin{align}
A_{LF}(x^{LF})  &  =%
%TCIMACRO{\dsum }%
%BeginExpansion
{\displaystyle\sum}
%EndExpansion
\frac{1}{n!}\int_{H_{m}^{(\pm)}}..\int_{H_{m}^{(\pm)}}a(p_{1}^{LF}%
,..p_{n}^{LF})e^{-i(\sum_{k}p_{k}^{LF})x^{LF}}\times\label{pre}\\
\times &  :A_{in}(p_{1}^{LF})..A_{in}(p_{n}^{LF}):dp_{1}^{LF}..dp_{n}%
^{LF}\nonumber\\
x^{LF}  &  =(x_{+},\mathbf{x}),\ p^{LF}=(p_{-},\mathbf{p}),\ dp^{LF}%
=\frac{dp_{-}}{2p_{-}}d\mathbf{p}\nonumber
\end{align}
where, just as in the GLZ representation before, the mass shell representation
of the LF projection involves integrations over the positive and negative mass
hyperboloid $H_{m}^{(\pm)}$ (creation/annihilation part). Note that the
coefficient functions $a$ and the ($\pm$ frequency ) creation/annihilation
parts of $A_{in}(p)$ in (\ref{pre}) remain the same as in (\ref{GLZ}), only
the parametrization of the mass shell is now specified in terms of $p^{LF}.$
By construction the covariance of this field is given in terms of the
mentioned 7-parametric subgroup, any other transformation would lead out of LF
and therefore destroy the representation (\ref{pre}).

Hence superficially the pointlike fielf formulation of the LF projection seems
to work, but there is a hitch \cite{Re}. The holographic projection in d=1+3
fails on composite fields even in the absence of interactions. The easiest way
to see this problem is to apply the above projection formula to the lowest
composite $:A^{2}(x):$ (notation as before)%
\begin{equation}
:A^{2}(x):_{LF}=\frac{1}{\left(  2\pi\right)  ^{3}}\int e^{i(p_{1}^{LF}%
+p_{2}^{LF})x_{LF}}:A(p_{1}^{LF})A(p_{2}^{LF}):dp_{1}^{LF}dp_{2}%
^{LF}\label{sqare}%
\end{equation}
The absence of any transverse damping (a constant $a(p_{1}^{LF},p_{2}^{LF})$)
brings about the appearance of the square of a meaningless transverse delta
function $\delta(\mathbf{x})^{2}$ in the two-point function and the associated
commutator. For the construction of the holographically projected algebras one
does not need the existence of a LF projection for each individual composite
field. But there is no theorem which secures the existence of even a single
interacting field for which the above prescription (11) works.

The by far safest and most systematic, but unfortunately computationally least
accessible approach to null-surface holography and its inversion is the
algebraic method in the AQFT setting of nets of local algebras. This
construction does not require the existence if pointlike lightfront limite.
Its procedure interms of subsequent intersections of wedge algebras with a
common LF was indicated at the beginning of this section and more details can
be found in \cite{S1}\cite{interface}. As mentioned on a previous occasion,
the inversion of a holographic projection is highly nonunique. Knowing the
local substructure on LF, one can reconstruct all $\mathcal{A}(W)^{\prime}s$
(without their local substructure!) with $H(W)\subset LF$, which falls short
of the full local net. An attractive method to supply the missing off LF
external data for holographic inversion is the relative modular positioning
which in the LF context amounts to knowing something about the relation
between the algebras of differently positioned LFs ("GPS" in AQFT)
\cite{interface}.

The field generators whose holographic projection caused the problem of the
ill-defined transverse part do not play any direct role. At the end of the day
one may of course ask for generating fields of the holographic projection.
There should be problem to construct these objects since a transverse extended
chiral algebra is very similar to a chiral algebra and for the latter one
knows how to extract generating fields from the net of operator algebras.

In algebraic terms, the absence of transverse vacuum fluctuations means that
the global lightfront algebra tensor factorizes under \textquotedblleft
transverse subdivisions\textquotedblright:
\begin{align}
\mathcal{A}_{\mathrm{LF}}(\mathbb{R}^{2}\times\mathbb{R}_{+})  &
\cong\mathcal{A}_{\mathrm{LF}}(R\times\mathbb{R}_{+})\otimes\mathcal{A}%
_{\mathrm{LF}}(R^{\prime}\times\mathbb{R}_{+})\\
\Omega &  =\Omega_{R}\otimes\Omega_{R^{\prime}}%
\end{align}
where $R\subset\mathbb{R}^{2}$ and $R^{\prime}=\mathbb{R}^{2}\setminus R$, and
this factorization is inherited by subalgebras associated with intervals
$I\subset\mathbb{R}_{+}$ in the lightlike direction (see below). Although a
detailed derivation of the localization structure on the horizon of the wedge
requires a substantial use of theorems about modular inclusions and
intersections (for which we refer to \cite{Sch1}\cite{Sch2}), the tensor
factorization of the horizon algebra relies only on the following structural
theorem in operator algebras:

\begin{theorem}
(Takesaki \cite{Tak}) Let $\left(  \mathcal{B},\Omega\right)  $ be a von
Neumann algebra with a cyclic and separating vector $\Omega$ and
$\Delta_{\mathcal{B}}^{it}$ its modular group. Let $\mathcal{A}\subset
\mathcal{B}$ be an inclusion of two von Neumann algebras such that the modular
group $Ad\Delta_{\mathcal{B}}^{it}$ leaves $\mathcal{A}$ invariant. Then the
modular objects of $\left(  \mathcal{B},\Omega\right)  $ restrict to those of
$\left(  \mathcal{A}e_{\mathcal{A}},\Omega\right)  $ where $e_{A}$ is the
projection $e_{\mathcal{A}}H=\overline{\mathcal{A}\Omega}$ as well as to those
of $\left(  \mathcal{C}e_{\mathcal{C}},\Omega\right)  $ with $\mathcal{C}%
=\mathcal{A}^{\prime}\cap\mathcal{B}$ the relative commutant of $\mathcal{A}$
in $\mathcal{B}$ and $e_{\mathcal{C}}H=\overline{\mathcal{C}\Omega}.$
Furthermore the algebra $\mathcal{A} \vee\mathcal{C}$ is unitarily equivalent
to the tensor product $\mathcal{A}\otimes\mathcal{C}$ in the tensor product
Hilbert space.
\end{theorem}

In the application to lightfront holography we choose $\mathcal{B}%
=\mathcal{A}(W)\equiv\mathcal{A}_{\mathrm{LF}}(\mathbb{R}^{2}\times
\mathbb{R}_{+})$. Its modular group is the Lorentz boost $\Lambda_{W}(-2\pi
t)$ which in the holographic projection becomes a dilation. The dilation
invariance of the algebra $\mathcal{A}=\mathcal{A}_{\mathrm{LF}}%
(A\times\mathbb{R}_{+})$ is geometrically obvious and hence the prerequisite
of the theorem concerning the modular group is met. The relative commutant is
$\mathcal{C}=\mathcal{A}_{\mathrm{LF}}(R^{\prime}\times\mathbb{R}_{+})$. The
lightlike nature of the subalgebra is absolutely crucial for the tensor factorization.

A particularly interesting situation is obtained for d=1+1 since in that case
the holographic projection has no transverse part and hence every bulk field
has a well defined chiral holographic projection and the conditions for an
inverse holography appear especially favorable. In d=1+1 there exists a
infinite family of so-called "factorizing models" \cite{Ba-Ka} (in the 70s and
80s also referred to a "integrable QFTs"). They are distinguished by the
presence of very simple generators of the wedge algebra $\mathcal{A}(W)$ which
instead of the ()commutation of creation/annihilation operators satisfy the so
called Zamolodchikov-Faddeev (Z-F) commutation relations \cite{ZZ} which in
the simplest case are of the form%

\begin{align}
\tilde{Z}(\theta)\tilde{Z}^{\ast}(\theta^{\prime})  &  =S_{2}(\theta
-\theta^{\prime})\tilde{Z}^{\ast}(\theta^{\prime})\tilde{Z}(\theta
)+\delta(\theta-\theta^{\prime})\label{Z-F}\\
\tilde{Z}(\theta)\tilde{Z}(\theta^{\prime})  &  =S_{2}(\theta^{\prime}%
-\theta)\tilde{Z}(\theta^{\prime})\tilde{Z}(\theta)\nonumber\\
Z(x)  &  =\frac{1}{\sqrt{2\pi}}\int(e^{ip(\theta)x(\chi)}\tilde{Z}%
(\theta)+h.c.)d\theta\nonumber
\end{align}
They deviate from the standard creation/annihilation operators by the
appearance of the $S_{2}$ function which has the consequence that the $Z(x)$
are not point-local, however they turn out to be still "wedge-local"
\cite{S2}\cite{S3}. They share with the standard creation/annihilation
operators that they create vacuum polarization free states from the vacuum.

In that case the expansion of the interacting field in terms of the Z-F
operators%
\begin{equation}
A(x)=%
%TCIMACRO{\dsum }%
%BeginExpansion
{\displaystyle\sum}
%EndExpansion
\frac{1}{n!}\int_{V_{m}^{\pm}}..\int_{V_{m}^{\pm}}a(p_{1},..p_{n})e^{i\sum
p_{k}x}:\tilde{Z}(p_{1})..\tilde{Z}(p_{n}):\frac{dp_{1}}{2p_{10}}%
..\frac{dp_{n}}{2p_{n0}} \label{form}%
\end{equation}

The coefficient functions of this expansion have an important additional
structural attribute: the crossing property \cite{foun}%
\begin{align}
a(p_{1},..p_{n}) &  =\left\langle 0|A(0)|p_{1},...p_{n}\right\rangle
^{in}\label{cro}\\
\left\langle 0|A(0)|p_{1},...p_{n}\right\rangle ^{in} &  =~^{out}\left\langle
-p_{k+1},..,-p_{n}|A(0)|p_{1},...p_{n}\right\rangle _{c.o}^{in}\nonumber
\end{align}
The various formfactors with different particle distributions between incoming
and outgoing particles are related to one analytic master function which may
be identified with the vacuum polarization component as in (\ref{cro}), where
the c.o stands for the omission of contractions between in and out momenta.
The factorizing models form an infinite family of nontrivial properly
renormalizable theories whose existence can be, for the first time in the
history of QFT, can be mathematically established \cite{Lech}; this is also
the first setting of interacting theories for which the constructive power of
LF holography can be convincingly demonstrated.

But again, as in the case of the GLZ series, one has not been able to control
the convergence of the formfactor series. The existence proof of these models
uses ideas from AQFT which avoid formfactor series such as (\ref{form}) so
that one cannot draw conclusions about their convergence. There are however
encouraging arguments that at least the holographically projected series can
be summed. In the case of the massive Ising QFT, which is a well-known
factorizing model, one can extract from its formfactor series another series
representation for the two-point function of its holographic projection on the
$x_{+}$-horizon ($x_{-}=0$). Of particular interest is an infinite series
which represents the anomalous dimension of the order parameter which is known
to be\footnote{The short distance dimension of the order parameter in the
massive Ising model is $1/8$ but only half of this value is seen in the
holographic projection.} $1/16.$ The mere fact that the holographic projection
reproduces this value is not surprising, but that this value results from
summing a nontrivial infinite series is astonishing\footnote{The calculation
has been done in the \textit{critical limit} which, although conceptually very
different from the holographic projection, leads to the same anomalous
dimensions. In contrast to the latter the critical limit yields a different
theory in its own reconstructed Hilbert spaces.} \cite{Ba-KaW}. In other
models for which one has series representations one also expects convergence
of the series for anomalous dimensions and more generally for the series
representing the holographic projection of (\ref{form}).

The holographic method also sheds light on the conceptually mysterious but
computational successful Zamolodchikov proposal to consider factorizing models
as resulting from suitably perturbed conformal models. In the Zamolodchikov
setting the conformal model is viewed as the universality zero mass limit of a
factorizing model. The concept of holography suggests to substitute the
universality class limit by lightray holography in order to gain conceptual
clarity. Whereas a zero mass universality class limit of a massive theory
comes with a different Hilbert space, the holographic lightray projection of a
massive 2-dimensional model lives in the same Hilbert space as the massive
model and moreover their localization concepts are algebraically related. This
makes factorizing models very useful objects in a "theoretical laboratory "
for the study for investigating in particular the idea of lightlike holography
for a long time to come.

Factorizing models in d=1+1 also show an interesting structural phenomenon of
holography with respect to the spin and statistics issue. Observable chiral
fields can only come from Boson fields in the bulk with integer dimensions
which are typically conserved currents or energy-stress tensors. Bosonic
fields with anomalous dimensions (as most fields in factorizing models) pass
to plektonic (nontrivial braid group representation) fields as a result of
interlinking of spin, statistics and dimension in chiral theories. One expects
of course that those plektonic chiral fields are precisely those which one
obtains from the representation theory of the local observables.

The physical motivation for holography is not very different from that of its
predecessor, the "lightcone quantization"; the main difference is the
conceptual and mathematical precision coming from local quantum physics.
Lightcone quantization was introduced during the 60s as a simplifying tool for
exploring QFT in the nonperturbative regime, especially for high energies. But
it did not quite achieve what it was introduced for, partly because of old
misunderstandings about the conceptual nature of QFT. Without a few comments
on the interrelation between lightcone quantization and lightfront holography
one would miss a unique chance to appreciate the gain of insight into the
inner workings of QFT; and since such comments according to the best of my
knowledge cannot be found in other publications they will be briefly presented here.

Its name "lightcone \textit{quantization}" suggests a description in terms of
a different theory\footnote{In approaches starting from different
quantizations, it is necessary to establish a relation between them, a problem
which was not adequately addressed.}, whereas in reality it only changes the
spacetime view of the given (already quantized) quantum matter in such a way
that certain physical aspects of interest are focussed on at the expense of
blanking out others. There was another terminology namely "going into the
$p\rightarrow\infty$ frame". In this setting it was clear that one looked for
a simplification in the theory, but as a result of the predominant
perturbative momentum space methods the importance of conceptual understanding
in terms of the causal locality principle was not yet appreciated.

Lightcone quantization (in a good-natured view) and lightfront holography
agree on pointlike fields i.e. they can be applied (with the same results) on
all free fields, on all interacting fields in d=1+1 and on certain
higher-dimensional interacting fields. A field as (\ref{sqare}) does not
permit an LF limit in any any setting. Since LQP views fields as
coordinatizations of local operator algebras, the fact that may exist only a
few (or no) interacting fields which admit a pointwise LF projection does not
limit the holographic method. As explained before holography can be defined
directly in terms of nets of spacetime-indexed operator algebras in the bulk
which have their horizon on the same null-surface. In other words holography
is a change of spacetime encoding of a given abstract algebraic substrate
which does not rely on finding suitable pointlike generators. Since conformal
theories are known to be pointlike generated and since holographic projections
turn out to be transverse extended chiral theories, one expects fields on
null-surfaces to be always pointlike generated. However there may not be a
direct path from generating fields in the bulk to those in the horizon. 

That holography was born on the umbilical cord of lightcone quantization as an
attempt to overcome its conceptual misunderstandings becomes evident if one
looks at the first papers in particular the one by 't Hooft's \cite{Ho} where
the presently used terminology appears for the first time. However the reader
is alerted that what is called in the literature "the holographic principle"
has little to do with holography in this and other articles based on QFT
\cite{Moret}\cite{interface}, The holography onto null-surfaces is a genuine
projection and not a faithful encoding. The existence of a faithful map of the
information in the bulk onto a null-screen (i.e. a correspondence) contradicts
the locality structure. 

The theorists dream for such complex theories as QFT with infinite degrees of
freedom is to decompose the original problem into a collection of simpler
problems. Indeed, the QFT of extended chiral fields which appears after the
holographic projection is much simpler than the bulk theory. But every
simplification in QFT has its prize; in the present case there is no unique
holographic inverse without invoking additional informations from outside the
LF. In terms of degrees of freedom, the lightfront holography amounts to a
thinning out of degrees of freedom. By adding the knowledge of how some
Poincar\'{e} generators act on the holographic projection or, which amounts to
the same, about the relative positioning of QFT on different $H(W)$ "screens",
one recovers the larger cardinality of degrees of freedom which is necessary
to have a physically viable bulk theory. 

This would be very different for correspondences between bulk and time-like
("branes") subalgebras as the AdS-CFT correspondence \cite{Du-Re}\cite{foun}
with the CFT brane at infinity. In that case there is no adjustment of
cardinality of degrees of freedom as in the holographic projection onto causal
or event horizons. Rather the cardinality (together with the symmetry group)
stays the same, which renders one side unphysical \cite{Sw}; either the lower
dimensional side develops physical pathologies which show up in causal
propagation problems and in problems in introducing temperature states, or in
the opposite direction, the higher dimensional theory is too "anemic" in order
to support particle physics as we know it. 

It is strange that the phase space degree of freedom issue of QFT over QM
which took so much effort to be understood in the 1960-90 has been lost in the
mealstrom of the milleniums particle theory thus showing that the dictum "that
many people cannot err" is not only wrong in extra scientific human activities.

Although holography as an extension of the old lightcone quantization is
foremost an instrument to make QFT more amenable to calculations and improve
its conceptual understanding, there has been a lot of interest in making it
also useful for problems of black-hole physics and cosmological horizons. It
is interesting to look at these problems from a particle physics viewpoint. In
the next subsection it will be shown that the Bondi-Metzner-Sachs symmetry is
a consequence of the symmetry gain in the holographic projection.

\subsection{The quantum origin of the Bondi-Metzner-Sachs symmetry}

The holographic projection onto the lightfront inherits a 7-parametrig
subgroup from the 10 parametric Poincar\'{e} group of the bulk. As a result of
the loss of transverse vacuum polarization and the fact that holography onto
the horizon leads to a transverse extended chiral QFT with infinitely many new
symmetries. In addition to the Diff(S$^{1}$) invariance in lightlike
direction, the compactification of the transverse plane (which is compatible
with the quantum mechanical delta function) extends the infinity preserving
E(2) symmetry to the SL(2,C) fractional action on the z,\={z} Riemann sphere
with z=x+iy. Both transformations together generate a very large symmetry
group which contains in particular $z,\bar{z}$ dependent Diff(S$^{1}$).

Even if one restricts to transformations which leave the vacuum invariant
(proper symmetries) there are still infinite parameters since the parameters
of the Moebius group can be functions of $z,\bar{z}$ in such a way that a
SL(2,C) transformation leads to a change of the Moebius parameters which is
consistent with the composition law of the Moebius group. The full symmetry
group of $z,\bar{z}$ dependent lighlike diffeomorphisms is gigantic. As we
will see the BMS group \cite{Sachs} results from this vacuum preserving
subgroup generated by $z,\bar{z}$ dependent Moebius-transformations by
imposing in addition the preservation of lightlike infinity which only leaves
$z,\bar{z}$ dependent transformation of the dilation-translation kind $a+bx.$

In the following we want to argue that this is not an accidental consequence
of our special choice of taking the horizon of a wedge as our null-surface,
but that it holds as well for the holographic projection on the upper horizon
of a double cone which is part of the mantle of a lower light cone shifted
upward so that its lower end is t=0 plane. The asymptotic situation envisaged
by Penrose \cite{Wald} results by moving the upper apex to timelike infinity
in which the mantle defines what one means by lightlike infinity in a Penrose sense.

In the case of conformal models one can try to compute generators for
double-cone holography by applying the appropriate conformal transformation to
convert the wedge into a double cone. The conformal map from the $x_{0}-x_{3}$
wedge $W$ to the radius=1 double cone $\mathcal{O}_{1}$ placed symmetrically
around the origin is%
\begin{align}
\mathcal{O}_{1} &  =\rho(W+\frac{1}{2}e_{3})-e_{3},\text{ }\rho(x)=-\frac
{x}{x^{2}}\\
W &  =\left\{  (x_{0},x_{\perp},x_{3})\ |~x_{3}>\left\vert x_{0}\right\vert
,x_{\perp}\in R^{2}\right\}  \nonumber
\end{align}
with $e_{3}$ being the unit vector in the 3-direction. Restricted to the
(upper) horizon $\partial W$ one obtains in terms of coordinates%
\begin{align}
\partial\mathcal{O}_{1} &  \ni(\tau,\vec{e}(1-\tau)),~\tau=\frac{t}%
{t+x_{\perp}^{2}+\frac{1}{4}},~\vec{e}=\frac{1}{x_{\perp}^{2}+\frac{1}{4}%
}(x_{\perp},\frac{1}{4}-x_{\perp}^{2})\\
&  where\text{ }\partial W=\left\{  (t,x_{\perp},t)\ |\ t>0,~x_{\perp}\in
R^{2}\right\}  \nonumber
\end{align}
If we use the unitary conformal transformation $\mathcal{A}(W)\rightarrow
\mathcal{A(O}_{1}\mathcal{)}$ not only on global generators for $\partial
\mathcal{A}(W)=\mathcal{A}(W)$ but also for their pointlike generating fields
$A_{LF}$ (\ref{LF}), we obtain the desired compact transverse proportionality
factor $\sim\delta(\vec{e}-\vec{e}^{\prime})$ replacing $\delta(x_{\perp
}-x_{\perp}^{\prime})$ from the fact that the t-independent relation between
$\vec{e}$ and $x_{\perp}$ is that of a stereographic projection of $S^{2}$ to
$R^{2}.$ The presence of this factor corroborates the absence of transverse
vacuum polarization in the above algebraic argument. The lightlike factor has
the expected qualitative behavior in terms of the variable $\tau$ and the
$W$-modular group $t\rightarrow e^{\lambda}t$ passes to the $\mathcal{O}_{1}$
modular automorphism
\begin{equation}
\tau\rightarrow\frac{-e^{-\lambda}(\tau+1)+1}{e^{-\lambda}(\tau+1)+1}%
\end{equation}
The transverse additive group passes via inverse stereographic transformation
to the transverse rotational group.

The generators for $\mathcal{A}(\partial\mathcal{O}_{1})$ are obtained by
conformal transforming the lightfront generators (\ref{LF}). In order to
notice that the full transverse symmetry is 6-parametric, one should realize
that already before the transformation the transverse quantum mechanics with
the fluctuationless vacuum state has a higher symmetry than just the
3-paramteric Euclidean symmetry of a plane. For this purpose it is helpful to
perform the stereographic projection to the Riemann sphere. The latter has the
6-parametric SL(2,C) group as its symmetry group and this brings immediately
to ones mind that this is related to the fractional action of the (covering of
the) Lorentz group on the space of unit vectors (or lightlike directions).
This action creates a conformal factor which, as a result of the additional
conformal factors arising from the conformal covariant lightlike variable, can
easily be compensated.

So no matter whether we study the holographic projection onto $\partial W$ or
$\partial\mathcal{O}_{1}~$we find the same symmetry acting on the
transverse$\times$lightlike coordinates ($z,\bar{z}$)$\times u$\footnote{Here
we pass to the cosmologically more costumary notation $u$ instead of $x_{+}.$%
}. The total dynamical symmetry groups is the infinite-parametric group of all
$z,\bar{z}$ dependent diffeomorphisms on the line extended by $z,\bar{z}$
automorphism of the Riemann sphere whereas the proper (vacuum preserving)
symmetry group is the $z,\bar{z}$ extended Moebius group of the circle. Here
we are interested in the ax+b subgroup of the Moebius group which acts on the
uncompactified lightray.

The z,\={z} dependence leads to the Bondi-Metzner-Sachs group\footnote{There
have been several attempts to relate the classical BMS group with quantum
physics \cite{Da}.}%

\begin{align}
u  &  \rightarrow F_{\Lambda}(z,\bar{z})(u+b(z,\bar{z}))\label{first}\\
(z,\bar{z})  &  \rightarrow U(\Lambda)(z,\bar{z}),~U(\Lambda)\in
SL(2,C)\nonumber
\end{align}
The group composition law $F_{\Lambda^{\prime}}(\Lambda(z,\bar{z}))F_{\Lambda
}(z,\bar{z})=F_{\Lambda^{\prime}\Lambda}(z,\bar{z})$ requires the
multiplicative factor to be of the form%
\begin{equation}
F_{\Lambda}(z,\bar{z})=\frac{1+\left\vert z\right\vert ^{2}}{\left\vert \alpha
z+\beta\right\vert ^{2}+\left\vert \gamma z+\delta\right\vert ^{2}}%
,~A(\Lambda)=\binom{\alpha,\beta}{\gamma,\delta}%
\end{equation}
Here the Greek letters stand for the SL(2,C) parametrization of the covering
of the Lorentz group whereas the functions $b(z,\bar{z})$ are from a function
space which is the closure of $C^{\infty}(z,\bar{z})$ functions on the Riemann
sphere in some topology.

Some more comments are in order. The first line (\ref{first}) is just a tiny
part of the infinite dimensional chiral conformal diffeomorphism group; in
fact it is the transverse extended infinity-fixing subgroup of the
vacuum-preserving Moebius-group. It receives its parametric enrichment from
the fact the two parameters $(F,b)$ (\ref{first}) of this $as+b$ Anosov group
vary with the compactified\footnote{In general the spatial coordinates of a
QFT cannot be compactified, but here the absence of transverse vacuum
fluctuations comes to one's help.} $z$-coordinates. The Lorentz group appears
here as the symmetry group of the Riemann sphere which results from the
compactified transverse coordinates. This enlargement draws on the absence of
transverse vacuum fluctuations.

The somewhat unexpected property is that the action of SL(2,C) on the large
function space contains in its linear part a copy of the semidirect product
action of the Lorentz group on the translations i.e. the infinite dimensional
BMS group contains the Poincar\'{e} group.

Our mission to demonstrate that the BMS groups is a subgroup of the much
bigger holographic group onto a lightlike horizon is now accomplished and for
more informations, especially on the position of the Poincar\'{e} inside the
BMS group, we refer to a comprehensive paper by Dappiaggi \cite{Da2} and the
literature cited therein.

It is well known that historically this infinite dimensional group arose in
general relativity as an asymptotic symmetry in asymptotically flat solutions
of the Einstein Hilbert equations \cite{Wald}. In the present quantum context
this group describes the vacuum preserving and infinity fixing part of a
larger symmetry group which comes with the holographic projection onto
horizons independent of the context in which the "null-screen" arises. At
first sight this is surprising because whereas one can envisage a nontrivial
action of the Poincar\`{e} group on the asymptotic Penrose screen, it is not
clear what this means in the case of the horizon of a finite double cone. What
does the action on a causal horizon which is not left invariant by Poincare
transformation mean?

For this problem we use some properties of a so-called \textit{split
inclusion} of a double cone in a slightly bigger double cone $\mathcal{A(D(}%
R\mathcal{)\subset}$ $\mathcal{A(D(}R+\Delta R\mathcal{)}$ which will be the
subject of a more detailed study in the last section. \ The important
consequence of a split inclusion for the present problem concerning the action
of a copy of the Poincar\'{e} group on the horizon of a double cone is that
there exists a canonical factorization of the Hilbert space $H=H_{1}\otimes
H_{2}$ such that $\mathcal{A(D(}R\mathcal{)}$ acts on the factor $H_{1}$ and
the causal disjoint of the bigger algebra $\mathcal{A(D(}R+\Delta
R\mathcal{)}^{\prime}$ (which is spatially separated from the smaller by a
security distance $\Delta R$) acts on $H_{2}.$ In this case it was shown that
there exists in a canonically distinguished representation of the Poincare
group which locally (and for small group elements so that the image stays
inside $\mathcal{D}(R)$) acts on operators in $\mathcal{D}(R)$ as the original
representation. This Poincare group acts in $H_{1}$ i.e. on the algebra
$B(H_{1})$ $\supset\mathcal{A(D(}R\mathcal{)}$ where the localization of
$B(H_{1})$ in $\mathcal{D}(R+\Delta R)$ is sharp in $\mathcal{D}(R\not )$ and
"fuzzy" in the surrounding sheet of size $\Delta R.$ With respect to this only
partially physical representation of the Poincare group the boundary horizon
$\partial\mathcal{D}(R+\Delta R)$ behaves like a Penrose screen and the
universal appearance of the BMS symmetry in holographic projections on
null-surfaces looses some of its mystery. In other words the split inclusion
of two double cones as above only describes the physical world inside the
smaller double cone. Beyond its boundaries the split Poincare group acts
mathematically correct as if the ring region and it boundary defined by the
mantle of the larger double cone would be the infinite remainder of the world
and its Penrose infinity. But this is an artifact of the split construction
since the action of the split Poincare group only coalesces with its physical
counterpart inside the smaller double cone; the outside region up to the
$R+\Delta R$ boundary is a fake part of the split universe on which, different
from the region inside $R,$ the split Poincar\'{e} group acts nongeometrically
\cite{Haag}. So at least the element of surprise of encountering a BMS-Penrose
situation already on a null-surface in finite spacetime has been removed; the
BMS group is a subgroup of the very big symmetry group which emerges after
holographically projecting matter onto null horizons\footnote{Since holography
is a process in which only the localization of quantum matter is changed
whereas the Hilbert space is maintained, the infinite parametric holographic
groups act also on the bulk, but this action is extremely fuzzy (non
geometric) and therefor not of physical interest.}. The physical relation with
the classical BMS work is limited to the infinitely far Penrose horizon.

The universal validity of the BMS symmetry on horizons raises the question
whether holography on finite lightlike surfaces as $\mathcal{D}(R),$ which are
null-surfaces but not lightfronts, continues to hold for massive i.e. not
conformal theories. It is well-known that a free massive QFT cannot have a
geometric acting modular group, the action inside the double cone must be
"fuzzy". In the previous section we argued that this results from the
geometric modular action on the horizon because the massive propagation from a
horizon into the bulk is "reverberating".

This problem has a classical analog: the propagation of characteristic data on
a null-surface into the bulk. The relevant formula (\ref{com}) which can be
found in a recent paper \cite{Re} works for the classical and for the
interaction free quantum case.

\subsection{Split property and entropic area law}

As well known, the Unruh effect \cite{Unruh} can be viewed as a thermal
manifestation of the causal localization which underlies QFT but is absent in
QM. The Hamiltonian is not the usual one associated with time translation in a
Minkowski inertial system, but rather (up to a multiple which depends on the
acceleration of the observer) the wedge-preserving Lorentz boost generator $K$
in a Rindler world. The vacuum restricted to the causally closed wedge region
is a KMS state at the modular temperature $2\pi$ associated to $K.$

It is believed that this is not just a mathematical discovery with the only
purpose to highlight some unusual conceptual and structural aspects of QFT
\cite{Bi-Wi}, but rather represents an in observational (in principle, but in
practice not directly accessible) effect, which an appropriately uniformly
accelerated thermal radiation counter will actually register. Its spacetime
basis is the fact that the uniform acceleration of a particle counter causally
confines the latter to a wedge-shaped spacetime region and at least
subjectively converts the original global relativistic "world time" into the
"Rindler time".

Whereas the observer's time on each of the uniformly acceleration orbits
inside a Rindler world is a concept of classical general relativity, the
thermal aspect comes from the modular localization properties of QFT. The
Unruh effect associated to the Rindler time on the different uniform
acceleration orbits different temperatures in a way which maintains the well
known (imaginary) time- temperature relation in passing from the abstract
modular time/temperature to that of the Unruh Gedankenexperiment.

Trading an inertial system with a uniformly accelerated one entails changing a
positive energy Hamiltonian (with the vacuum being the bottom state) with a
boost Hamiltonian whose energy spectrum is two-sided. In this way the vacuum
becomes a thermal state in which the zero energy only refers to a mean value
in the middle. So what appears as a small change in the spacetime situation
has a conceptually large repercussion on the local quantum level. In the end
it leads to a change in the association between the Hamiltonian and the
measurement hardware as well as a change in the perception of the global
vacuum state.

Although these radical conceptual changes lead to numerical modifications
which remain way below observational accessibility, they are unavoidable
consequences of QFT, a theory which has remained the most successful and
comprehensive description of our material nature. Therefore they require the
utmost conceptual and philosophical attention. Historically the thermal
manifestation of localization has been first observed in curved spacetime QFT
in the presence of event horizons, which in contrast to the fleeting causal
horizons in flat spacetime, have an observer-independent status defined in
terms of intrinsic properties of spacetime.

The Unruh situation is a special case of a more general setting of "modular
localization" \cite{S2}\cite{BGL}\cite{Fa-Sc}\cite{MSY} which describes the
position of the dense subspace in terms of domains of unbounded operators $S$.
These domains are determined by in terms of the unitary representation of the
Poincar\'{e} group but for knowing the operator itself one needs to know
dynamical aspects of a QFT which leads to those localized states. The full
S-operator is defined in the algebraic setting \cite{Haag} of QFT as%

\begin{equation}
S_{\mathcal{O}}A\Omega=A^{\ast}\Omega,\text{ }A\in\mathcal{A(O})
\end{equation}

where $\mathcal{A(O})$ is the operator algebra localized in $\mathcal{O}%
$\footnote{Without loss of generality we assume that the localization regions
are causally closed i.e. $\mathcal{O=O}^{\prime\prime}$ where one upper dash
denotes the causal disjoint.} and the state is (in most applications of QFT)
the vacuum state. Although there is no operator which maps all bounded
operators of the algebra of all operators into the adjoint, for certain kind
of operator algebras which includes the spacetime localized algebras of QFT
there does exist an unique unbounded such operator. The existence of an
uniquely defined "Tomita" operator $S$ in QFT is guarantied for a large class
of states, including the vacuum $\Omega$. The necessary and sufficient
condition is that the operator algebra $\mathcal{A(O})$ acts on the state
$\Omega$ in a cyclic and separating way (or shorter that ($\mathcal{A(O}%
),\Omega$) is in "standard position"). Unless specified otherwise $\Omega~$in
the sequel denotes the vacuum.

It turns out that this unbounded closed antilinear and involutive operator
encodes the \textit{causal completion} $\mathcal{O}^{\prime\prime}$ in its
domain in the sense that the change of localization region is precisely
mirrored in the change of $domS_{\mathcal{O}}$ in Hilbert space. \ $S$ has a
polar decomposition%
\begin{equation}
S=J\Delta^{\frac{1}{2}}%
\end{equation}
where the modular group $Ad\Delta^{i\tau}$ is an object of a "kinematical"
rather than dynamical nature\footnote{In the case of $\mathcal{O=}$ $W$ the
$\Delta^{it}$ is (up to a scaling factor) the W-preserving Lorentz boost. In a
system of particles obeying the mass gap hypothesis one conventionally regards
the particle spectrum "kinematical" (given) and considers as dynamical only
those properties which depend on the interaction between those particles.}
whereas the anti-unitary $J$ is "dynamic", since it depends on the scattering
matrix (see below).

Another remarkable property which follows from its definition is the
involutivity on its domain $S^{2}\subset\mathbf{1}$ and the \ "transparency"
of its domain in the sense of $ranS=domS=dom\Delta^{\frac{1}{2}}.$ Hence the
dynamics is encoded in a re-shuffling of vectors inside $domS.$ It turns out
that the global vacuum $\Omega,$ defined as the lowest state in a theory with
positivity of the (global) energy, after restriction to the subalgebra
$\mathcal{A(O})$ becomes a thermal KMS state with respect the modular
Hamiltonian $\Delta^{i\tau}=e^{-i\tau K}$

In fact in the case of a wedge algebra the dynamical content of the modular
reflection $J_{W}$ is given by\footnote{This observation of the author was
obtained by rewriting the TCP covariance of the Smatrix in an asymptotically
complete QFT and can be found in \cite{S3} (and earlier references therein).}%
\begin{equation}
J_{W}=J_{0}S_{scat}%
\end{equation}
where $J_{0}$ is the TCP-related modular reflection of the incoming free field
and $S_{scat}$ is the associated S-matrix. Its appearance as a relative
(between interacting and free algebras) modular invariant is surprising and
has powerful consequences of which some will be mentioned later. Although the
domain of the Tomita $S-$operator for subwedge algebras allows no direct
characterization in terms of the Poincar\'{e} group for subwedge regions,
these domains can be build up from intersections of $S_{Wedge}$ domains.

The general modular situation is more abstract than its illustration in the
context of the Unruh Gedankenexperiment since the generic modular Hamiltonian
is not associated with any spacetime diffeomorphism; it describes a "fuzzy"
movement which only respects the causal boundaries but is somewhat nonlocal
inside. In this case the existence of such a Hamiltonian is nevertheless of
structural value, since it allows to give a mathematically precise quantum
physical description of the locally restricted vacuum as a KMS state
associated with the intrinsically determined modular Hamiltonian. As argued in
the previous section, it is a fact that the holographic projections onto
horizons will convert the fuzzy acting modular Hamiltonians associated with
causally closed bulk subregions into geometric acting "surface Hamiltonians"
\cite{S2} which is represented by a the generator of a dilation. In this way
the possible loss of certain bulk symmetries is more than compensated for by
the gain of infinitely many new symmetries after the projection.

None of the above properties holds in QM where the only localization is the
probabilistic Born localization for which the space of $\mathcal{O}-$localized
wave functions at a fixed time is described by a projector $P_{\mathcal{O}}$
which results from the spectral resolution of the position operator. In that
case the vacuum simply factorizes, so that Born localization does not lead to
a new entangled state; in particular any kind of entanglement from
inside/outside localization factorization can never be of a thermal kind
unless the global state was already thermal from the beginning. It can be
shown that this factorization continues to hold for the ground states of
nonrelativistic finite density zero temperature matter.

Wigner tried to adapt the Born localization to the relativistic realm and
realized to his dismay\footnote{As a result of what he considered as a serious
flaw, Wigner maintained a critical distance towards QFT in the later part of
his life.} that this probability aspect is inconsistent with covariance
\cite{N-W} and reference dependent. The covariant modular localization, which
underlies the formalism of relativistic QFT, deals with dense subspaces which
cannot be described by projectors. Nevertheless the Born-Newton-Wigner
localization plays a crucial role in scattering theory a fact which results
from the fortunate circumstance that the correlation between asymptotically
timelike separated Born-Newton-Wigner (BNW) localized events is covariant
which leads to the consistency between covariance and the probability concept
on the level of the S-matrix and the scattering cross-section.

The fundamental difference between BNW and modular localization is reflected
in a radically different nature of local algebras $P_{\mathcal{O}%
}B(H)P_{\mathcal{O}}=B(P_{\mathcal{O}}H)$ in QM and $\mathcal{A(O})$ in QFT.
Localization in QM always ends up with the algebra of all bounded operators of
a smaller Hilbert space, more precisely on a factor space\footnote{In order to
facilitate the comparison with QFT we take the Fock space formulation of QM.}
$B(H)=B(P_{\mathcal{O}}H)\otimes B((\mathbf{1}-P_{\mathcal{O}})H)$ which
corresponds to the spatial decomposition $H=P_{\mathcal{O}}H\otimes
(\mathbf{1}-P_{\mathcal{O}})H.$

Whereas the total algebra in QFT is still of the form $B(H),$ localized
operator algebras in QFT are of hyperfinite type III$_{1}$ factor algebra (the
"monad" of QFT) in the classification of Connes, which constitutes a
refinement of the original classification by Murray and von Neumann. For the
sake of brevity (and also to avoid an unnecessary "shock" and awe effect with
the reader) we will call this algebra a monad, implying with this notation
that all localized $\mathcal{A(O})$ in QFT\footnote{The time slice property
which holds in all physically relevant theories states that the algebra of a
region is identical to that of the causally closed region. Therefore there is
no loss of generality from assuming that all regions are causally complete.}
are isomorphic copies of the monad which in turn is not isomorphic to the
quantum mechanical algebras from bipartite splits done with Born localization.
Again the monad $\mathcal{A(O})$ commutes with causal disjoint (which happens
to be equal to its commutant) $\mathcal{A(O}^{\prime})=\mathcal{A(O})^{\prime
}$ and both algebras span $B(H)=\mathcal{A(O})\vee\mathcal{A(O}^{\prime})$ but
this generation of the full algebra from its commuting parts cannot be brought
into the form of a tensor product. In fact the whole conceptual framework of
QM breaks down: the reduction of a pure state on $B(H)$ gives an impure state
which is not described in terms of a density matrix, and with the absence of
the tensor factorization for the bipartite partition of $B(H)$ the rug is
pulled out from under the standard usualsetting of entanglement.

The usual method of calculation ignores these structural properties and
proceeds with a formal "as if" calculation based on a quantum mechanical
tensor factorization. The unavoidable ultraviolet divergencies are interpreted
as a shortcoming if not inconsistency of QFT which needs a high energy
modification coming from a future quantum theory of gravity \cite{BKLS}.

The view taken in the sequel of this paper is more prosaic namely that the
divergence is not a shortcoming of QFT but rather the manifestation of the
radically different nature of local monad algebras which do not tensor
factorize under causal bipartite subdivision so that the prerequisite of the
entanglement-setting for quantum information theory is violated. The split
property permits to approximate a causally localized monad by a sequence of
quantum mechanical algebras whose localization region is slightly larger by a
distance $\Delta R$ which surrounds the localization region (a lighlike
"sheet" of thickness $\Delta R$ instead of a horizon) and serves as a kind of
attenuation distance for the otherwise infinite vacuum fluctuations at the
sharp boundary horizon. This split approximation permits a quantitative
description of the leading behavior in $\Delta R\rightarrow0.$ Of particular
interest is the limiting behavior of the entropy, which, as will be argued
below, is a logarithmically corrected area law in which the sheet distance
$\Delta R$ enters in such a way as to lead to a dimensionless result.

The split property of QFT leads to a very different kind of factorization as
that resulting from spatial bipartite factorization\footnote{In QM the ground
state factorizes whereas the "split vacuum" (or finite energy particle
states), as a result of vacuum polarization, is always a Gibbs-like thermal
state with an infinite number of particles.} in QM; whereas in the former case
the global vacuum does not factorize but rather becomes a Gibbs state upon
restriction to the split local algebra (the "split vacuum"), the quantum
mechanical vacuum factorizes. Although the splitting recovers a tensor
factorization, the factorising split vacuum is a "hot vacuum polarization
soup" a property which is often (misleadingly) attributed to the global
vacuum. 

The implementation of the analog of the thermodynamic limit KMS state by a
sequence of Gibbs states on box quantized (type I) algebras is precisely the
role of the split construction, in which a KMS state resulting from the
modular restriction of the vacuum to sharply localized algebra $\mathcal{A(O}%
)$ is approximated by a sequence of type I split algebras\ with \textit{fuzzy}
boundaries. The Gibbs states are obtained from the restriction of the vacuum
to the split algebra with the fuzzy boundary. The approximation can be done
either from the outside ("funnel") or from the inside (outside ---%
%TCIMACRO{\TEXTsymbol{>} }%
%BeginExpansion
$>$
%EndExpansion
inside under causal complement transformation); in a canonical way from the
restriction of the vacuum to by the on the states on local monads the funnel
limit to approximate trick is to approximate the desired region by a sequence
of regions with "fuzzy" boundaries and to realize that this process only leads
to vacuum polarization \textit{within} the fuzzy boundary. The observables in
this localization-caused thermal behavior depend on the thickness $\Delta R$
of the boundary and similar to the entropy of heat bath systems, which can
only be explicitly computed in the thermodynamic limit, one expects that the
localization entropy at best be computed for $\Delta R\rightarrow0$. We will
now illustrate this split setting in the special context which will be of
interest in the subsequent derivation of localization entropy.

For the following computation of localization entropy we will use the notation
of the previous section. Hence let $\mathcal{O=D(}R\mathcal{)}$ be the double
cone which results from the causal completion of a ball of radius $R$ around
the origin and consider a slightly bigger concentric ball with associated
double cone $\mathcal{D}(R+\Delta R).$ Then the inclusion $\left\{
\mathcal{A(D}(R))\subset\mathcal{A(D}(R+\Delta R)),\Omega\right\}  $ is called
standard if $\left\{  \mathcal{A(D}(R)),\Omega\right\}  ,\left\{
\mathcal{A(D}(R+\Delta R)),\Omega\right\}  $ and $\left\{  \mathcal{A(D}%
(R+\Delta R))\cap\mathcal{A(D}(R))^{\prime},\Omega\right\}  $ are standard.
For standard inclusions one defines the split property as the existence of an
intermediate quantum mechanical type I algebra $\mathcal{N}$ i.e.%

\begin{align}
&  \mathcal{A(D}(R))\subset\mathcal{N}\subset\mathcal{A(D}(R+\Delta
R))\label{inc}\\
&  \mathcal{A}(ring)\equiv\mathcal{A(D}(R))^{\prime}\cap\mathcal{A(D}(R+\Delta
R)),~\nonumber\\
&  \mathcal{N=A(D}(R))\vee J_{ring}\mathcal{A(D}(R))J_{ring}\nonumber
\end{align}

If the standard inclusion is split, there are infinitely many intermediate
type I algebras and among those there is a "canonical" one\footnote{The
formula for the canonical $\mathcal{N}$ does not by itself secure the type I
factor property; it only defines a canonical split, if the inclusion is split
to begin with .}, which is uniquely determined by the modular data and given
by the formula in the third line \cite{Do-Lo}. The tensor factorization
$B(H)=\mathcal{N\otimes N}^{\prime},~H=P_{N}H\mathcal{\otimes(}%
1-P_{\mathcal{N}}\mathcal{)}H=H_{1}\otimes H_{2}$ where $P_{\mathcal{N}}$ is
the projection onto the subspace $\mathcal{N}\Omega,$ together with the
inclusions gives the desired tensor split%
\begin{equation}
\mathcal{A(D}(R))\vee\mathcal{A(D}(R+\Delta R))^{\prime}\simeq\mathcal{A(D}%
(R))\otimes\mathcal{A(D}(R+\Delta R))^{\prime}\label{split}%
\end{equation}
i.e. the statement that the operator algebra generated by the smaller and the
commutant of the larger is isomorphic to their tensor product. But in contrast
to the quantum mechanical factorization, the vacuum state does not factor but
rather is highly entangled and leads, upon reduction to the factor algebra
$\mathcal{N}$, to a thermal Gibbs state associated with Hamiltonian determined
by the modular data of the split situation \cite{Do-Lo}.

We previously mentioned the useful analogy between the "funnel" limit $\Delta
R\rightarrow0$ in the thermal setting of local algebras of QFT and the
thermodynamic limit $V\rightarrow\infty$ in the setting of QM. At this point
it is important to be reminded of the fact that, although the Born localized
algebras of ground state QM are always type I and this continues to be the
case for box-quantized Gibbs systems, the thermodynamic limits of Gibbs
systems are hyperfinite type III$_{1}$ algebras on which the Gibbs state
changed into a more singular KMS state which cannot be described in terms of a
density matrix. In fact the radical change of the type I algebra to a type
III$_{1}$ monad algebra in the thermodynamic limit is directly related to the
volume divergence.

In both cases one approaches a monad in a KMS state by a sequence of Gibbs
states on quantum mechanical type I algebras. The main difference is one in
physical interpretation; in one case one approximates a KMS state on a global
monad (which is interpreted as a global algebra) by an increasing sequence of
type I algebras, whereas in the other case the approximating type I sequence
is shrinking for $\Delta R\rightarrow0$ towards the monad which is interpreted
as $\mathcal{A(D}(R))$. This analogy suggests to expect that the divergences
in both cases become identical after an appropriate reparametrization which
takes care of the different geometric aspects. In the rest of this section we
will collect supporting arguments for the following statement:

\textbf{Statement} \textit{Global heat bath systems in the thermodynamic limit
and local split systems in the limit of vanishing split distance have the same
divergent entropy factors after replacing the dimensionless volume by the
logarithmically modified dimensionless area }
\begin{equation}
V_{n-1}\left(  kT\right)  ^{n-1}|_{T=T_{\operatorname{mod}}}\simeq\left(
\frac{R}{\Delta R}\right)  ^{n-2}ln\left(  \frac{R}{\Delta R}\right)
\label{st}%
\end{equation}
\textit{where }$V_{n-1}$\textit{ is the standard dimensionful volume factor
which is made dimensionless with the Boltzmann factor and on the right hand
side appears the already mentioned dimensionless area factor}. \textit{The
modifies area behavior should be seen as a "lightlike volume factor" with the
transverse directions giving the area factor and the lightlike direction (of
the light-like sheet of thickness }$\Delta R$\textit{) contributing the
logarithmically parametrized missing length factor.}

Here the left hand side is the dimensionless (as a result of the kT factors)
"volume" divergence which appears in the thermodynamic limit of an
n-dimenional QFT. The right hand side represents the divergence factors in the
in the split limit $\Delta R\rightarrow0.$ This relation for n=4 implies in
particular that the localization entropy follows an logarithmically corrected
area behavior.

One may view this statement as indicating an "inverse Unruh effect" i.e. the
idea that \textit{a thermodynamic limit state on a global operator algebra can
be viewed as a localized subsystem of a larger system for which a global pure
state (e.g. a global vacuum) has been restricted to the subsystem}. The
necessity to fix the heat bath temperature at a particular value is no
surprise in view of the existence of the Hawking temperature. For n=2 and
conformal invariant theories one can show that the two systems are isomorphic,
with the unitary equivalence corresponding to a conformal transformation
\cite{S1}. For higher dimensions the presence of the logarithmic factor is the
contribution from the lightlike direction in a light sheet. 

In any case, the idea that the Hawking temperature originates from the
localization thermality of QFT and the (logarithmically corrected) entropical
area law is part of a totally different setting, namely a still unknown
quantum gravity, seems to be untenable. Localization entropy is without doubt
a concept of QFT and the role of curvature is to convert observer-dependent
"fleeting" causal horizons into the intrinsically positioned event horizons of
QFT in curved spacetime.

The support for the statement comes from three different directions.

\begin{enumerate}
\item The singularity of the entropy for $\Delta R\rightarrow0$ and that of a
dimensionless "partial charge" in a similar geometric situation are
consequences of the same mechanism of vacuum polarization.

\item Both situations, that of a thermodynamic open system and that of a
modular localized operator algebra are mathematically identical, namely
hyperfinite type III$_{1}$ operator algebras in a KMS state, only the physical
parametrization in terms of approximating quantum mechanical type I algebras
is different

\item For n=2 the relation (\ref{st}) expresses the inverse Unruh effect in
which the heat bath length factor passes to the logarithmic short distance
behavior from localization on a lightlike line. This is a consequence of the
existence of a conformal transformation connecting the two systems.
\end{enumerate}

As an historical interlude it is interesting to mention that Heisenberg's
discovery of vacuum polarization led to a behavior which is very similar the
one in the statement. He found that if one integrates the zero component of
the conserved current of a charged free field over a finite spatial region of
radius R, the so defined "partial charge" diverges, which brings us to the
first point. In the modern QFT setting it is possible to control the strength
of such a divergence in terms specially prepared localizing test functions. A
finite partial charge with a vacuum polarization cloud within a ring of
thickness $\Delta R$ is is defined in terms of the following
\textit{dimensionless} operator%
\begin{align}
&  Q(f_{R,\Delta R},g_{T})=\int j_{0}(\mathbf{x},t)f_{R,\Delta R}%
(\mathbf{x})g_{T}(t)d\mathbf{x}dt\label{partial}\\
&  \left\Vert Q(f_{R,\Delta R},g_{T})\Omega\right\Vert \equiv F(R,\Delta
R)\nonumber
\end{align}
where the spatial smearing is in terms of a test function which is equal to
one inside a sphere of radius R and zero outside $R+\Delta R$ with a smooth
transition in between and $g_{T}$ is a finite support $\left[  -T,T\right]  $
interpolation of the delta function. As a result of charge conservation such
expressions converge for $R\rightarrow\infty$ to the global charge either
weakly \cite{Sw} or (of one relates T with R appropriately) even strongly on a
dense set of states \cite{Requ}.

For an estimate of the vacuum polarization one would like to study the limit
of $\Delta R\rightarrow0$ for fixed $R$ of $F(R,\Delta R)$ (\ref{partial}). As
expected, the computation for the chiral conformal case shows a
logarithmically divergent partial charge, whereas for additional spatial
dimension a factor of the dimensionless area $\frac{area}{\left(  \Delta
R\right)  ^{n-2}}$ appears. The calculation of the two-point function of the
smeared zero component of the current produces the naively expected result,
apart from a logarithmic factor. Note that the smearing in timelike (or
lightlike) direction in addition to the spatial $f_{R,\Delta R}$ smearing is
essential in order to obtain a finite operator, without it there would be no
logarithmic divergent contribution. The rotational symmetry as well as the
dimensionless nature of a partial charge would suggest a behavior as the right
side of (\ref{st}). This time the calculation can be explicitly done in terms
of two-point functions of a concrete operator.

The localization entropy on the other hand is not a computable property of a
specific dimensionless testfunction-smeared operator-valued distribution as
the partial charge, but rather a manifestation of the vacuum polarization
cloud residing in the lightlike sheet around the horizon $\partial
\mathcal{A(D)}$ of algebra $\mathcal{A(D)}\footnote{Since the vacuum
polarization in both cases of the flucuation of the partial charge and of the
entropy of the localized bulk matter is caused by the same localization
mechanism, the identical behavior for $\Delta R\rightarrow0$ should be of no
surprise.}.$ What links both cases is the universality of vacuum polarization
caused by localization, independent of whether the localization occurs through
testfunction smearing leading to dimensionless operators, or through
localization of an entire algebra which takes one to the dimensionless
entropy. Such analogies are however no replacement for computations,
especially in cases in which the concepts entering the computation are more
interesting than the result itself.

The assumption of conformal invariance simplifies the computation and confirms
the area behavior which one could directly have obtained by a dimensional
argument. The dimensionless logarithm (which is the only vacuum polarization
contribution in two-dimensional theories) escapes such arguments and therefore
the chiral case requires an explicit calculation which will be done directly
for the inverse chiral Unruh effect in the later part of this section.

An analog behavior of $F$ to (\ref{st}) may be taken as an indication that the
vacuum polarization leads to a universal divergent behavior if the attenuation
length of the vacuum polarization cloud is made to shrink $\Delta
R\rightarrow0,$ independent of whether the vacuum polarization is caused by
the algebraic split property for localized algebras or whether it is coming
from individual operators whose formal limit describe (dimensionless) partial charges.

A mathematical proof that the localization property for a finite split
localization entropy which in the $\Delta R\rightarrow0$ limit behaves as
claimed (\ref{st}) amounts to a control of the density matrix $\rho
_{split}\sim e^{-K_{split}}$ and the entropy is the von Neumann entropy of
this Gibbs state (there is a corresponding density matrix on the commutant
$\mathcal{N}^{\prime}$ which leads to the same entropy). In the present state
of QFT technology this is a hopeless task. As will be shown in the sequel
there are special circumstances related to the knowledge about chiral theories
which allow to do this for n=2.

The important aspect of the above split inclusion of two double cones is that
the vacuum looks only different from what it was before the split in a
ring-like region whose associated algebra is the relative commutant of the
smaller within that localized in the bigger double cone. $\mathcal{N}$ is the
canonically associated type I algebra in terms of which there is tensor
factorization as in (\ref{split}). The relative commutant in the second line
is of special interest since geometrically it describes the finite shell
region (or rather its causal completion) in which we expect the vacuum
polarization to be localized. The restriction of the vacuum to $\mathcal{N}$
is a density matrix state $\rho_{split}$ since the algebra is quantum
mechanical; In principle one could compute the entropy exactly on the basis of
the modular data for $\mathcal{N}$, but in practice this is (as in the analog
case of the thermodynamic limit) in the present state of knowledge about
modular theory only possible in the funnel limit $\Delta R\rightarrow0.$

The \textit{split tensor factorization} leads to a notion of entanglement
which is still distinctively different from the information theoretical
entanglement which one encounters in QM \cite{interface}. A bipartite tensor
factorization associated with the quantum mechanical Born localization creates
(upon restriction to one tensor factor) a density matrix which is not of the
thermal kind unless the global state was thermal to begin with. In other words
quantum mechanical spatial bipartite partitions create information theoretic
entanglement but generally do not lead to thermal KMS properties for the
restricted states. Thermal manifestations of localization and hence of
bipartite splitting is only possible in QFT; physically because one needs
localization-caused vacuum polarization and mathematically, since the
(sharply) localized algebras are monads whose properties are radically
different from quantum mechanical type I factor algebras (monads have no
density matrix states) even though they can be approximated by type I algebras.

In no way does modular localization and splitting create a real temperature
which is associated with the physical Hamiltonian and the physical time. But
there are zillions of other "Hamiltonians" within the same QFT model, i.e.
Hermitian operators with respect to which the vacuum is not a state at the
lower end of a one-sided spectrum, but for which the spectrum is two-sided.
Modular theory selects a particular Hamiltonian with two-sided spectrum and
what lends physical importance to this otherwise mathematical construction is
the fact that the selection is inexorably coupled to localization which is the
most important (and most subtle) property of particle physics. The modular
groups associated with such Hamiltonians are automorphisms of the localized
algebras which map of the bulk matter in a geometrically fuzzy way maintaining
however the causal boundaries (horizons); that they represent a diffeomorphism
is the exception and happens for massive theories in Minkowski spacetime only
in the case of the Rindler wedge leading to the Unruh Gankenexperiment.

Even though effects related to modular localization theory will probably never
be directly observational accessible (since they are orders of magnitudes
smaller than quantum mechanical entanglement effects historically related to
Schroedinger's cat the violation of Bell's inequality,...), their importance
cannot be overestimated if it comes to the problem of nonperturbative
classification and constructions of interacting models of QFT. A more detailed
discussion of this point will be given in the next section.

A much more solid situation, in which horizons are defined by the system and
not by the observer, results from event horizons in curved space time. The
best known and historically first example of such a horizon is the
Schwarzschild solution. In fact the thermal aspects of localization have been
first observed by Hawking \cite{Haw} while performing calculations on free
scalar quantum fields in the Schwarzschild metric.

In the present work we will avoid curved spacetime because it is our intention
to convince the reader that many conjectured properties which arose in curved
spacetime QFT and for which the presence of gravity was thought to be
essential are in fact preempted by more abstract mathematical and conceptual
properties in flat spacetime QFT. The basic difference between the flat and
the curved spacetime situation is that the modular Hamiltonian associated with
a thermal description of localized quantum matter bears no relation to the
physical Hamiltonian associated with a time translation in an the inertial
system, whereas in case of event horizons as in the Schwarzschild spacetime
the modular Hamiltonian of ($\mathcal{A(O}_{out})$,$\Omega_{H.H.}$) is
identical to that related to the timelike Killing symmetry; here
$\mathcal{O}_{out}$ denotes the part outside the black hole and $\Omega
_{H.H.}$ is the Hartle-Hawking state on the Kruskal extension of the
Schwarzschild spacetime \cite{HHstate}.

We now return to the issue of localization entropy in QFT in the context of
the before mentioned standard split inclusion of double cone.

Let us start with the case of a two-dimensional conformal QFT in which case
the double cone is a two-dimensional spacetime region consisting of the
forward and backward causal shadow of a spatial line segment of length $R$ at
$t=0$ sitting inside a bigger causal shadow region obtained by extending the
baseline on both sides by $\Delta R.$ As a result of the assumed conformal
invariance of the theory, the canonical split algebra inherits the
covariances, and hence the entropy of the canonical split algebra can only be
a function of the cross ratio of the 4 points characterizing the split
inclusion%
\begin{align}
S &  =-tr\rho ln\rho=f(\frac{\left(  d-a\right)  \left(  c-b\right)  }{\left(
b-a\right)  \left(  d-c\right)  })\label{cross}\\
with~a &  <b<c<d=-R-\Delta R<-R<R<R+\Delta R\nonumber
\end{align}
where for conceptual clarity we wrote the formula for the conformal invariant
ratio in case of generic position of 4 points. For chiral theories the
dependence of the entropy on the cross ratio of 4 points on the lightray
expresses the fact that the entropy is a conformal invariant. In comparison
with higher dimensions one does not need this generality, the symmetric case
written in the last line (\ref{cross}) is sufficient. Our main interest is to
determine the leading behavior of $f$ in the limit $\Delta R\rightarrow0$
which is the analog of the thermodynamic limit $V\rightarrow\infty$ for heat
bath thermal systems.

The asymptotic estimate for $\Delta R\rightarrow0$ can be carried out with an
algebraic version of the \textit{replica trick\footnote{The replica trick is
well-known in mathematical work on solid state problems, including the
calculation of entropy \cite{solid}. }} which uses the cyclic orbifold
construction in \cite{Lo-Fe}. First we write the entropy in the form
\begin{equation}
S=-\frac{d}{dn}tr\rho^{n}|_{n=1},~\rho\in M_{can}\subset\mathcal{A}(R+\Delta
R)
\end{equation}
Then one uses again the split property, this time to map the n-fold tensor
product of $\mathcal{A}(L+\Delta L)$ into the algebra of the line
(conveniently done in the compact $S^{1}$) with the help of the $n^{th}$ root
function $\sqrt[n]{z}.$ The part which is invariant under the cyclic
permutation of the n tensor factors defines the algebraic version \cite{Lo-Fe}
of the replica trick. The transformation properties under the higher analogs
of the Moebius group are now given in terms of the following subgroup of
DiffS$^{1}$ written formally as%
\begin{align}
&  \sqrt[n]{\frac{\alpha z^{n}+\beta}{\bar{\beta}z^{n}+\bar{\alpha}}},~L_{\pm
n}^{\prime}=\frac{1}{n}L_{\pm n},~L_{0}^{\prime}=L_{0}+\frac{n^{2}-1}{24n}c\\
&  \dim_{\min}=\frac{n^{2}-1}{24n}c\nonumber
\end{align}
where the first line is the natural embedding of the n-fold covering of Moeb
in DiffS$^{1}$and the corresponding formula for the generators in terms of the
Virasoro generators. As a consequence the minimal $L_{0}^{\prime}$\ value
(spin, anomalous dimension) is the one in the second line. With this
additional information coming from representation theory we are able to
determine at least the singular behavior of $f$ for coalescing points
$b\rightarrow a,$ $d\rightarrow c$%
\begin{equation}
S_{sing}=-lim_{n\rightarrow1}\frac{d}{dn}\left[  \frac{(d-a)(c-b)}%
{(b-a)(d-c)}\right]  ^{\frac{n^{2}-1}{24n}}=\frac{c}{12}ln\frac{(d-a)(c-b)}%
{(b-a)(d-c)}\label{chiral}%
\end{equation}
Since the function is only defined at integer n, one needs to invoke Carlson's theorem.

The resulting entropy formula in the singular limit reads%
\begin{equation}
S_{sing}=\frac{c}{12}\ln\frac{(d-a)(c-b)}{(b-a)(d-c)}=\frac{c}{12}%
ln\frac{R(R+\Delta R)}{\left(  \Lambda R\right)  ^{2}} \label{ent}%
\end{equation}
where $c$ in typical cases is the Virasoro constant (which appears also in the
chiral holographic lightray projection).

This result was previously \cite{S1} obtained through establishing the
validity of the "inverse Unruh effect" for chiral theories. This is a theorem
stating that for a conformal QFT on a line, the KMS state obtained by
restricting the vacuum to the algebra of an interval is unitarily equivalent
to a global heat bath temperature state at a certain (geometry-dependent)
value of the temperature \cite{BoYn}\cite{Sc-Wi}. The transformation turns out
to be a conformal transformation \cite{S3} which carries the L
(one-dimensional volume) divergence into a $ln\varepsilon^{-1}$ factor with
$\varepsilon\sim e^{-L}$ and hence describes the same leading logarithmic
divergence as the more detailed argument using the replica trick.

The existence of the inverse Unruh effect in chiral theories is an explicit
demonstration that the above logarithmic divergence is nothing but the
conventional volume (here length) factor conformal transformed into an
exponential transformation of the affine length into the scaling group
parametrization. This corresponds to a direct transformation of the large
distance thermodynamic limit to a "funnel limit" in which a sequence of
quantum mechanical type I algebras obtained from splitting converge towards
the monad algebra of the smaller region for shrinking split
distance\footnote{The funnel approximation can also be made from the inside.}.

Although the inverse Unruh effect is apparently restricted to chiral theories,
the intriguing analogy of the heat bath entropy with the localization entropy
continues to exert itself. Below it will be argued that the localization
entropy in the n-dimensional case diverges for $\Delta R\rightarrow0,$ with
$\Delta R$ the splitting distance, as%
\begin{equation}
E\overset{\Delta R\rightarrow0}{=}C(n)\frac{R^{n-2}}{\left(  \Delta R\right)
^{n-2}}cln\frac{R^{2}}{\left(  \Delta R\right)  ^{2}}\label{en}%
\end{equation}
where we combined all numerical constants into $C(n)$ except the Virasoro $c$
which in the present context has the interpretation of a parameter
corresponding to the holographically projected matter. It is the only
recollection on the bulk quantum matter of this otherwise universal asymptotic behavior.

Compared to the chiral models, which can be controlled quite elegantly with
the replica method, the question of higher dimensional localization entropy
looks more involved. Neither the replica method is applicable nor a higher
dimensional inverse conformal Unruh effect seems to be available. For
conformal theories one uses dimensional arguments. The rotation symmetry
together with the dimensionless of the entropy requires to multiply the two
dimensional logarithmic with the dimensionless area factor $\left(  \frac
{R}{\Delta R}\right)  ^{2}$ from the two transverse directions in d=1+3..

It should be possible to check this for higher dimensional conformal free
fields; this would remove the last veil of mystery between localization and
heat bath thermal aspects in QFT. The validation of such a universality would
amount to a significant step in the understanding of local quantum physics
i.e. in the QFT beyond the Lagrangian setting. It would support the idea that
the approximation of a monad by tensor factor algebras is, after a
"kinematical" adjustment for the different spacetime situations\footnote{In
the heat bath case the monad is indexed by the entire Minkowski spacetime but
the thermal representation is unitarily inequivalent to the vacuum
representation. In the case of localized bulk matter the spacetime indexing is
a causally complete subregion and the state is the reduced vacuum.
\par
{}} as in (\ref{st}). The clarification about the presence/absence of this
universality would also be important for ideas on the interrelation of
thermodynamics, geometry and gravitation and support ideas\footnote{There are
two ongoing attempts at QG, the one favored by the already existing strong
relation with statistical mechanis and the other one invoking noncommutative
spacetime from confronting the idea of uncertainty relations with classical
aspects of black holes. There is no position operator and hence no Heisenberg
uncertainty relationin QFT, but the thermal aspect of localization shows that
the size $\varepsilon$ of the vacuum polarization cloud (the region of
fuzziness) is related to the entropy (or mean energy) content.} proposed by
Jacobson \cite{Jac}. 

The important aspect of the split inclusion of two double cones is that the
vacuum looks only different from what it was before the split in a ring-like
region (\ref{inc}) whose associated algebra is the relative commutant of the
smaller within the bigger double cone. The relative commutant in the second
line is of special interest since geometrically it describes the finite shell
region (or rather its causal completion) in which we expect the vacuum
polarization to be localized in that ring. The restriction of the vacuum to
$\mathcal{N}$ is a density matrix state $\rho_{split}$ on $\mathcal{N}$\ the
subalgebra $\mathcal{A(D}(R))\subset\mathcal{N}$ is indistinguishable from the
vacuum expectation values, so that only if tested with operators in the ring
region the "split vacuum" differs from the original vacuum. The split entropy
is the von Neumann entropy of the mixed state $\rho_{split}$; as in the
inside/outside tensor-factorization in QM the density matrix of the opposite
factor $\mathcal{N}^{\prime}$ leads to the same entropy.

In principle one could compute the entropy exactly on the basis of the modular
data for $\mathcal{N}^{\prime}$ but practically this is (as in the analog case
of the thermodynamic limit) only possible in the funnel limit $\Delta
R\rightarrow0.$ The logarithmic factor is the camouflaged third length factor
which hides the complete analogy of the area law with the thermodynamic volume
factor. For the derivation of the formula for the canonical $\mathcal{N}%
^{\prime}$ see \cite{Do-Lo}.

The resulting formula (\ref{en}) has a clear derivation in the conformal case
because besides the length $R$ which determines the hypersurface "area"
$R^{n-2}$ the only other dimension carrying parameter is $\Delta R$ so that
the entropy is given by (\ref{ent}) with a kinematical proportionality factor
$C(n)$ which depends on the spacetime dimension but unlike c is independent on
the quantum matter. It is believed that massive matter does not change the
leading behavior for $\Delta R\rightarrow0.$ The dynamically nontrivial part
of the argument is the derivation of the chiral entropy in terms of the
conformally invariant cross ratio of 4 points; the remaining steps consists
basically of symmetry and dimensional arguments.

The derivation of (\ref{chiral}) based on the split inclusion is preceded by
arguments using functional integral representations and momentum space
cutoffs. They naturally inherit all the conceptual problems of the use of
functional integrals in QFT\footnote{Whereas functional integrals have a solid
mathematical status for standard problems of QM, they are limited to
(Euclidean) free field actions and to fields whose short distance properties
are not worse than those of free fields (superrenormalizable interactions).
They are in particular not valid for fields with anomalous short distance
dimension which includes all factorising (integrable) QFTs.}. On a formal
level the introduction of a momentum space cutoff corresponds to the
introduction of the $\Delta R$ split distance. But whereas the latter is
clearly defined construct within a local theory, the introduction of the
former wrecks the locality of the theory in an uncontrolled way. This
situation has to be distinguished from the intermediate use of
cutoffs/regulators in renormalization theory as a calculational device to be
removed at the end of the calculation. In the present context the use of
momentum space cutoffs masks the fact that the spatial bipartite division does
not lead to a factorization which is the prerequisite for the standard notion
of entanglement. by the necessity of cutting off momentum space integrals in
order to avoid infinities. 

Momentum space cutoffs in QFT have severe conceptual problems of their own.
First there is no argument that a Euclidean cutoff functional integral is
still associated with a quantum theory; in fact in none of the explicitely
constructed two-dimensional factorising models it has been possible to
introduce a mathematically controllable  momentum space cutoff. A second
problem is that one cannot be sure whether the divergence only indicates an
avoidable inappropriate argument which leads through metaphoric intermediate
steps to a intrinsically consistent and correct result (example: the use of
cutoffs and regulators in renormalization theory), or whether behind the
divergence there is a universal structural limiting behavior of a physical
quantity, as is the case for the split entropy law (\ref{en}). The
localization entropy only depends on geometric (localization) data, it is
independent numerical factors which relate the modular Hamiltonian with that
of an observer (e.g. the Unruh acceleration).

Perhaps the oldest entropy calculation for a bipartite situation in QFT is
that in \cite{BKLS} which was done in the aftermath of Bekenstein's
conjecture. Apart from the logarithmic factor, the result (\ref{en}) agrees
with the present result. The comparison with Bekenstein's entropy formula is
obtained by a re-interpretation of a classical gravitational area law in the
thermal black hole setting \cite{BKLS}. However the suggestion that QFT needs
for reasons of conceptual-mathematical  consistenty a cutoff at the Planck
length is not correct, the consistency of models of QFT has nothing to do with it.

The computation starts from the (as we know now) incorrect assumption of a
splitless tensor factorization (as it is possible in QM but not in QFT
\cite{interface}) and pays the prize in form of a momentum space divergence
which, as customary in many textbook QFT, is dumped into a momentum space
cutoff. The area dependence (without the logarithmic factor) is then inferred
by dimensional reasoning from the cutoff dependence. The nature of the
localized vacuum polarization near the horizon remains somewhat hidden and the
interpretation of the entropy in terms of a momentum space cutoff goes into
the wrong physical direction. One would be hard pressed to conclude from such
computations that the vacuum polarizations which cause this phenomenon are
localized near the horizon and that the momentum space cutoff is related to
the size of the polarization sheet.

But whereas a momentum space cutoff limits the validity of the theory and
modifies it in an uncontrollable way, the split method shows that although the
localization entropy of sharply localized quantum matter is really infinite,
there is no reason to modify the model by a momentum space cutoff\footnote{The
idea in favor of a finite black hole entropy cannot be based on arguments in
\cite{BKLS} since localization entropy can be concistently defined without
momentum space cutoffs.}. The splitting procedure is a manipulation on the
given (unsplit) system which, just as the holographic projection keeps the
Hilbert space. Both methods are consistent with the local covariance principle.

In QFT there exists no position operator and hence there is no Heisenberg
uncertainty relation. On can view the relation between the split required
$\varepsilon$-"collar" region around a localized algebra and the vacuum
polarization induced entropy/energy (\ref{en}) as the QFT analog of the the
quantum mechanical $x-p$ uncertainty relations. A frame independent position
operator, as is well known does not exist in Poincar\'{e} covariant QFT; this
makes attempts to enforce noncommutativity/nonlocality based in position
operators questionable on conceptual grounds.

\section{Concluding remarks, outlook}

In this note we analyzed the quantum aspects of two problems which had their
origin in classical general relativity, namely the BMS symmetry in
asymptotically flat curved spacetime theories, and the Bekenstein area law for
event horizons in classical geometric field theories of the Einstein-Hilbert type.

In the first case we found that the symmetry gain in holographic projections
leads to an infinite symmetry group whose vacuum and infinity preserving part
is the BMS group. This infinite parametric group contains the Poincar\'{e}
group, but only if the nullsurface is that of conformal infinity in the sense
of Penrose does the Poincar\'{e} group agree with the physical one. The
difficulty is to understand the interpretation of the Poincar\'{e} subgroup of
BMS in case that the null-surface is not the Penrose infinite lightlike
boundary. This somewhat paradoxical situation was interpreted by observing
that the split situation creates a fictitious continuation of the smaller
algebra into the ring-like slice region in which the larger boundary appears
as an infinitely remote lightlike Penrose "screen". We emphasized that the
gain in infinite dimensional symmetry through holographic projection is
inexorably related to the thinning out of degrees of freedom and the
impossibility to reconstruct the bulk from only the intrinsic properties of
the projection.

QFT leads to a logarithmically corrected are law in which the logarithmic
factor is the only singular factor in case of chiral theories on a light-line.
The \textit{chiral inverse Unruh effect} explains this factor as resulting
from a conformal transformation applied to the length factor (i.e. the
one-dimensional volume factor). This observation brings the volume
proportional heat bath entropy and the logarithmically modified dimensionless
area law into a much closer relation than hitherto imagined. The area aspect
is the only one shared with the Bekenstein formula; whereas QFT produces a
one-parametric family of $\Delta R$-dependent entropies, the Bekenstein
formula achieves the dimensionlessness of entropy with the help of
dimensionful constants from classical gravity. has no such parameter. Whereas
the Hawking radiation is fully explained in terms of QFT in a Schwarzschild
CST, the Bekenstein formula can only be reconciled with QFT if instead of
ordinary quantum matter it refers to an unknown non-localizable gravitational
form of quantum matter.

In both cases the principles of local quantum physics led to unexpected
properties which were previously overlooked and which, at least in the case of
the area proportionality of entropy near a horizon, add new aspects to the
ideas around the Bekenstein entropy and its possible connection with the still
illusive quantum gravity. It is important to have a good understanding of
horizons and entropy caused by vacuum polarization near horizons first within
the setting of QFT before embarking on the more ambitious program of quantum gravity.

Fundamental to all problems addressed in this article is the principle of
causal localization, whose intrinsic mathematical formalization is the theory
of modular localization and the mathematical (Tomita-Takesaki) modular theory
of operator algebras. This is the best way to take care of the holistic
aspects of local quantum physics.

Finally it may be helpful to remind the reader that the QM-QFT antagonism with
respect to localization \cite{interface} and the resulting dichotomy between
information-theoretical and thermal entanglement is the influential
observation made in 1964 by Haag and Swieca \cite{Ha-Sw}\cite{Sw} that there
exists a fundamental difference between the cardinalities of phase space
degrees of freedom in QM and QFT; thus destroying once and for all the once
popular idea that QFT may be viewed as some kind of relativistic QM. Whereas,
as everybody learns in a course on QM, a finite cell in phase space only
accommodates a finite number of degrees of freedom, the phase space occupation
in QFT is infinite, albeit a quite "tame infinity", namely the phase space
densities form compact sets, later refined to nuclear sets \cite{Bu-Wi}. This
increase of degrees of freedom is a characteristic property of QFT, and
although the aim of Haag and Swieca to understand the asymptotic completeness
property of scattering theory in terms of phasespace degrees of freedom
cardinality remains open up to date, the enormous fertility of this idea in
the structural understanding of QFT is outside of any doubt.

Models with transnuclear phase space behavior may still mathematically exist
as local theories, but they have a series of pathological properties which
make particle physicists dismiss them as unphysical. In particular their
thermal behavior has either a limiting temperature (Hagedorn temperature) or
thermal states do not exist. Parallel to the thermal changes there is a change
in the causal shadow picture in that the causal dependency region contains
many more degrees of freedom than those which got there by propagation.

It is easy to provide illustrations: the AdS-CFT correspondence leads from a
standard AdS QFT to an overpopulated CFT. For example a free field on
AdS$_{5}$ leads to a generalized conformal free field on the compactified
$M_{4}$ with a Kallen-Lehmann spectral function which fulfills a power law in
the invariant mass \cite{Du-Re}. It is comforting to know that this never
happens in the holographic projection of localized bulk matter on its horizon.
Writing the LF fields in terms of their intrinsic degrees of freedom on which
only a 7-parametric subgroup of the Poincare group acts, there are lesser
degrees of freedom on LF than in the bulk. This is a blessing because it makes
holography a powerful constructive tool of QFT.

Although from a mathematical viewpoint there is no big difference between
causal and event horizons, there are severe additional conceptual problems in
passing from QFT to QFT in CST. One such problem is the question of what
reference state to take, since there is no distinguished replacement for the
vacuum state in generic CST. Since the holographic projection leads to a much
simpler transverse extended chiral theory, it may be more natural to discuss
the problem first in the holographic projection with its much simper state
structure, and than to extend this state to the bulk. Such a procedure was
proposed and illustrated in \cite{DMP}\cite{Moret}.

It would be desirable and add credibility to the ongoing discussions about the
still elusive QG, if it would take place with the full knowledge of the
holistic structure of QFT which is in sharp contrast to (relativistic)
QM\footnote{A relativistic QM or direct "particle interaction theory" (DPI) is
a relativistic QM (no vacuum polarization) which fulfills all properties which
one can formulate in terms of pacrticles without using interpolating fields
(invariant S-matrix, cluster properties) \cite{interface}.}. Holography on
horizons, the absence of tensor factorization under splitless causal bipartite
dissection, the possibility to characterize a full QFT including the action of
the Poincar\'{e} symmetry from the relative positioning of a finite number of
monads and, last not least, in agreement with the local covariance
principle\footnote{The implementation of this principle requires to study all
theories on different but isometric manifolds simultaneously in order to
construct a QFT on a particular QFT \cite{BFV}.}, all these properties
illustrate the holistic aspects of QFT which have no counterpart in QM. As
stressed by Hollands and Wald \cite{Ho-Wa} these aspects become important in
applications of QFT to cosmology e.g. in estimates of energy densities in
cosmic reference states which replace the Minkowski vacuum; simply adding up
energy modes as in QM violates this holistic property of QFT.

\end{document}